\documentclass[10pt,journal,compsoc]{IEEEtran}

\ifCLASSOPTIONcompsoc
  \usepackage[nocompress]{cite}
\else
  \usepackage{cite}
\fi
\ifCLASSINFOpdf
\else
\fi
\usepackage{amsmath,amsfonts}
\usepackage{algorithmic}
\usepackage[ruled,linesnumbered]{algorithm2e}
\usepackage{graphicx}
\usepackage{textcomp}
\usepackage{xcolor}
\usepackage{comment}
\usepackage{caption}
\usepackage{subcaption}

\usepackage{url}

\begin{document}
%
\title{FSpGEMM: An OpenCL-based HPC Framework for Accelerating General Sparse Matrix-Matrix Multiplication on FPGAs}
%
%
%
%

\author{Erfan Bank Tavakoli,
        Michael Riera,
        Masudul Hassan Quraishi,
        Fengbo Ren
\IEEEcompsocitemizethanks{\IEEEcompsocthanksitem E. Bank Tavakoli, M. Riera, M. Quraishi, F. Ren are with the School of Computing and Augmented Intelligence, Arizona State University, Tempe,
AZ 85281.\protect\\
E-mail:  ebanktav@asu.edu, mriera1@asu.edu, mquraish@asu.edu, renfengbo@asu.edu.
}
}

\IEEEtitleabstractindextext{%
\begin{abstract}
General sparse matrix-matrix multiplication (SpGEMM) is an integral part of many scientific computing, high-performance computing (HPC), and graph analytic applications. This paper presents a new compressed sparse vector (CSV) format for representing sparse matrices and FSpGEMM, an OpenCL-based HPC framework for accelerating general sparse matrix-matrix multiplication on FPGAs. The proposed FSpGEMM framework includes an FPGA kernel implementing a throughput-optimized hardware architecture based on Gustavson's algorithm and a host program implementing pre-processing functions for converting input matrices to the CSV format tailored for the proposed architecture. FSpGEMM utilizes a new buffering scheme tailored to Gustavson's algorithm. We compare FSpGEMM implemented on an Intel Arria 10 GX FPGA development board with Intel Math Kernel Library (MKL) implemented on an Intel Xeon E5-2637 CPU and cuSPARSE on an NVIDIA GTX TITAN X GPU, respectively, for multiplying a set of sparse matrices selected from SuiteSparse Matrix Collection. The experiment results show that the proposed FSpGEMM solution achieves on average 4.9$\times$ and 1.7$\times$ higher performance with 31.9$\times$ and 13.1$\times$ lower energy consumption per SpGEMM computation than the CPU and GPU implementations, respectively.

\end{abstract}

}

\maketitle





 


\section{Introduction}
\IEEEPARstart{G}eneral sparse matrix-matrix multiplication (SpGEMM) plays an important role in many scientific and high-performance computing (HPC) applications, such as multigrid interpolation/restriction \cite{briggs2000multigrid}, Schur complement methods in hybrid linear solvers \cite{yamazaki2010techniques}, colored intersection searching \cite{kaplan2006colored}, finite element simulations based on domain decomposition \cite{hapla2012use}, molecular dynamics \cite{itoh1995order}, and interior point methods \cite{karypis1994parallel}. SpGEMM is also widely used in graph analytic applications, such as graph contraction \cite{gilbert2008unified}, recursive formulations of all-pairs shortest-paths algorithms \cite{d2007r}, peer pressure clustering \cite{shah2007interactive}, cycle detection \cite{yuster2004detecting}, Markov clustering \cite{van2000graph}, triangle counting \cite{azad2015parallel}, and matching algorithms \cite{rabin1989maximum}.

However, the existing CPU- and GPU-based solutions to SpGEMM have limited computational performance due to two main reasons. First, an efficient implementation of SpGEMM algorithms (\textit{e.g.}, inner product, outer product, and Gustavson's \cite{gustavson1978two} algorithms) requires a tailored buffering scheme for storing and accessing intermediate results. The deep memory hierarchy and the fixed architecture of CPUs and GPUs make the implementation of such algorithm-tailored buffering schemes hardly efficient. Therefore, computing SpGEMM on CPUs and GPUs suffers from limited performance due to poor cache locality and frequent off-chip memory access for buffering intermediate results. Second, the complex operations (\textit{e.g.}, sparse vector addition) of SpGEMM algorithms impose strong loop-carried data dependency. The legacy architecture of CPUs and GPUs is able to exploit spatial parallelism but lacks the temporal/pipeline parallelism necessary for resolving such loop-carried data dependency, resulting in large loop initiation intervals that further limits the performance. Additionally, CPU- and GPU- based solutions to SpGEMM for HPC applications suffer from very high energy consumption due to long runtime (\textit{i.e.} low performance) and high power consumption (\textit{e.g.}, 120W-300W). The high energy consumption of CPUs and GPUs inevitably causes thermal and environmental conservation concerns in HPC and data center computing.

With the recent adoption of FPGAs in data centers \cite{escobar2015suitability} as an emerging accelerator, FPGA computing offers an alternative solution to accelerating SpGEMM. An FPGA is a farm of configurable hardware resources whose functionality and interconnection can be redefined at runtime by programming its configuration memory\cite{kuon2008fpga}. A state-of-the-art FPGA carries an enormous amount of fine- and coarse-grained logic, computation, memory, and I/O resources. Upon the reconfiguration of these resources, an FPGA can implement any custom hardware architecture tailed to a specific algorithm, often achieving performance and/or energy efficiency gains over CPU and GPU counterparts \cite{dorrance2014scalable,rafique2014communication,farhadi2019novel,bank2019polar}. The fine-grained logic resources and the abundant on-chip memory and register resources on FPGA devices can be configured to implement a custom buffering scheme tailored for a specific SpGEMM algorithm to allow efficient on-chip storage of intermediate results and data movement among processing elements (PEs) to avoid off-chip memory access\cite{tessier2001reconfigurable,woods2017fpga}. In addition, the reconfigurable resources on FPGA devices can compose not only spatial but also temporal/pipeline parallelism both at a fine granularity and on a massive scale to resolve the complex loop-carried data dependency that exists in SpGEMM algorithms to minimize loop initiation intervals for improved performance \cite{markovic2012dsp,biookaghazadeh2018fpgas}. Furthermore, while providing higher performance for executing high-dependency algorithms, FPGAs generally have lower power consumption than CPUs and GPUs, which directly translates into lower energy consumption per task (higher energy efficiency).
So far, there has been limited work on accelerating SpGEMM on FPGAs \cite{lin2013design,jamro2014algorithms}. 

Existing work all adopts the inner product algorithm \cite{srivastava2020matraptor} in their implementations. The inner product algorithm attempts to compute all zero and nonzero output values. In the case of sparse matrices, there are a considerable amount of computations that result in a zero output, consuming clock cycles that can be spared with domain knowledge embedded into the architecture. Additionally, the dot product operation between a row and a column of the input matrices requires index matching, which further contributes to the overhead of the SpGEMM algorithm. Therefore, the inner product algorithm is not suitable for SpGEMM.

In this paper, we propose a new compressed sparse vector (CSV) format for representing sparse matrices and FSpGEMM, an OpenCL-based HPC framework for accelerating SpGEMM on FPGAs. The proposed FSpGEMM framework consists of an FPGA kernel implementing a throughput-optimized hardware architecture and a host program implementing pre-processing functions for converting input matrices to the CSV format tailored for the proposed architecture. Different from prior work, we adopt Gustavson's algorithm to avoid zero output computation and reduce the synchronization overhead of computing partial products. Benefiting from the hardware flexibility of FPGAs, we propose a custom buffering scheme tailored to Gustavson's algorithm to improve the reuse of input matrices, thus largely reduce the amount of off-chip memory access. Additionally, we co-design FSpGEMM with the CSV format to transform the memory access pattern of input matrices from irregular to regular, which improves the memory bandwidth utilization and eliminates unnecessary memory stall. Overall, such synergies between the buffering scheme and Gustavson's algorithm as well as between the CSV format and FSpGEMM as a result of the co-design methodology significantly improve the computational performance.
The contributions of this paper are summarized as follows. 
\begin{itemize}
    \item We propose a new CSV format for representing and storing sparse matrices in a format that is co-designed with FSpGEMM for reducing unnecessary off-chip memory access and improving performance.  
    \item We propose an end-to-end OpenCL-based HPC framework for accelerating SpGEMM on FPGAs, consisting of an energy-efficient and deeply pipelined FPGA kernel for accelerating the Gustavson's algorithm and a host program implementing pre-processing functions for converting and/or storing raw input matrix files into the CSV format required for the proposed architecture.
    \item We propose a parameterized and scalable FPGA hardware architecture and an analytical method for determining the optimized architectural parameters for maximizing the run-time performance of SpGEMM given an arbitrary FPGA board specifications.
    \item We propose a new buffering scheme to improve the data reuse of Gustavson's algorithm and reduce the amount of off-chip memory access, which improves the overall computational performance.
    \item The proposed CSV format and framework can be potentially applied to all host-accelerator computing models with the SpGEMM computation offloaded to accelerator devices with customizable hardware (\textit{e.g.}, an FPGA or ASIC).
    \item We evaluate the performance and energy efficiency of FSpGEMM based on an Intel Stratix 10 GX FPGA development board and compare it with Intel Math Kernel Library (MKL) and cuSPARSE, the state-of-the-art libraries for SpGEMM, running on an Intel Xeon E5-2637 CPU and an NVIDIA GTX TITAN X GPU, respectively, for multiplying the poisson3Da, 2cubes\_sphere, filter3D, cage12, scircuit, mac\_econ\_fwd500, offshore, and webbase-1M matrices selected from SuiteSparse Matrix Collection \cite{davis2011university}. The experiment results show that the proposed FSpGEMM solution has on average 4.9$\times$ and 1.7$\times$ higher performance with 31.9$\times$ and 13.1$\times$ lower energy consumption per SpGEMM computation than the CPU and GPU implementations, respectively.
\end{itemize}

The rest of this paper is organized as follows. Section \ref{sec:back} provides background on different SpGEMM algorithms and summarizes the related work on accelerating linear algebra kernels on FPGAs as well as the state-of-the-art implementations of SpGEMM on CPUs, GPUs, FPGAs, and ASICs. Section \ref{sec:idea} elaborates the proposed CSV format. Section \ref{sec:design} describes the proposed framework's buffering scheme, hardware architecture, and the host program. The evaluation of FSpGEMM is presented in section \ref{sec:eval}. Finally, the paper is concluded in Section \ref{sec:con}.

\section{Background and Related Work}\label{sec:back}
\subsection{Sparse Matrix Formats}
Figure \ref{fig:format}) shows a compressed representation of a sparse matrix using Compressed Sparse Row (CSR) format \cite{bulucc2009parallel}. In this format, nonzero values of the sparse matrix are laid out in the row-major orientation in the off-chip memory. The CSR format stores a sparse matrix using three arrays $V$, $COL\_INDEX$, and $ROW\_PTR$ representing nonzero values and column index of the nonzero elements, and the pointer to the first nonzero element in the first two arrays, respectively.

In Compressed Sparse Column (CSC) \cite{bulucc2009parallel} format, the nonzero elements are stored in the column-major orientation using three arrays $V$, $ROW\_INDEX$, and $COL\_PTR$ for nonzero values, row index, and the pointer to the start of each column.

These data formats are not tailored to a SpGEMM algorithm or hardware architecture. Hence, they are not efficient for specific methods or hardware designs, which leads to a huge performance loss. Differently, the new CSV format makes the input data access in the proposed buffering scheme regular when utilizing the CSR or CSC formats.

\subsection{SpGEMM Algorithms}
There are three main methods for computing SpGEMM: inner product, outer product, and Gustavson's method. The differences among the three methods are twofold. First, these methods require different data formats for the input matrices to acquire contiguous access to off-chip memory. Second, some methods avoid wasted computations by calculating the nonzero elements of the output matrix.

The inner product algorithm \cite{srivastava2020matraptor} computes all the elements of the output matrix, including zero elements. In SpGEMM, most of the output elements are nonzeros. Additionally, computing each element involves a dot product operation, including index matching and multiply-accumulate (MAC). Jamro et al. \cite{jamro2014algorithms} identifies index matching as a hardware-expensive and the most time-consuming operation of the inner product method. Therefore, the inner product algorithm inevitably causes its implementations to summer from both performance and energy consumption overheads.

The outer product algorithm \cite{srivastava2020matraptor} performs an outer product operation between a column (row) of the first input matrix and row (column) of the second input matrix. The result of each outer product operation is a large partial sum matrix with the same dimensions as the input matrices. The number of partial sums produced is equal to the number of rows of the input matrices that are often large. Therefore, buffering and accessing partial sums requires off-chip memory access that incurs long access latency and consumes high energy. Moreover, the addition of large partial sums suffers from synchronization overhead. Consequently, the outer product algorithm suffers from undesired performance and energy consumption overhead.

\begin{figure}[t!]
\centering
\includegraphics[width=0.8\linewidth]{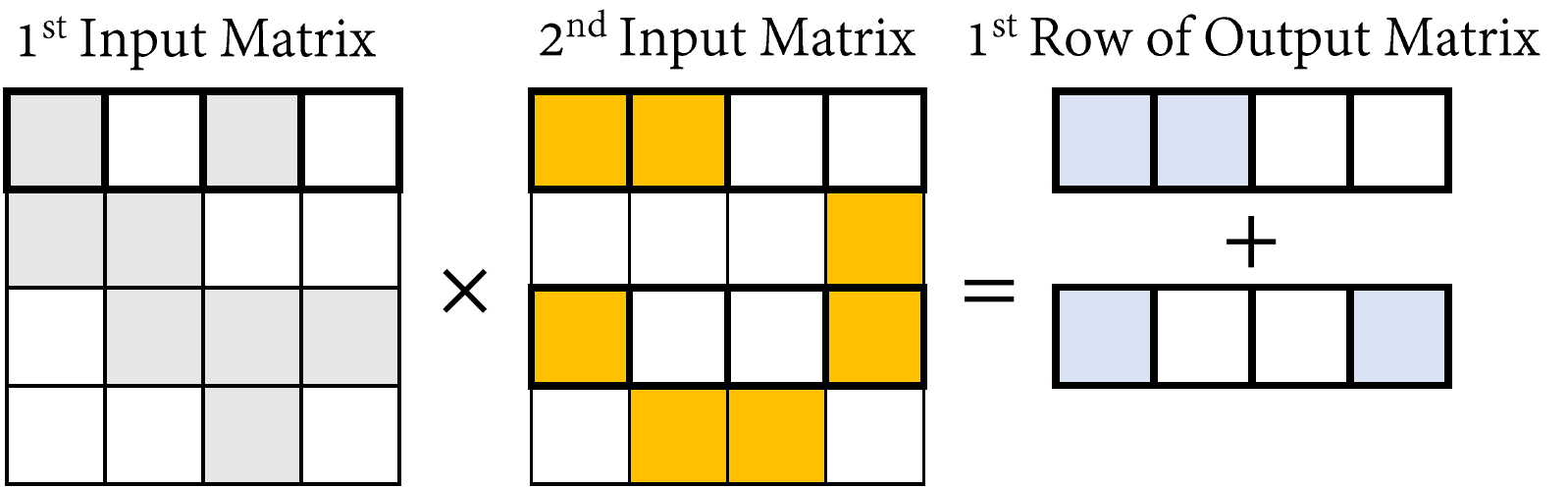}
\caption{An example for calculating the first row of the output matrix for SpGEMM using the row-wise Gustavson's method. The colored elements represent nonzero values, and the thick borders highlight the rows involved in the computation.}
\label{fig:gus}
\end{figure}

Figure \ref{fig:gus} illustrates the row-wise Gustavson's method for multiplying two sparse matrices ($A \times B$). In this method, each non-zero element in a row of the first input matrix (\textit{e.g.}, $A(i,j)$ where $i$ and $j$ are the row and column indices, respectively) is multiplied by all non-zero elements of the corresponding row of the second input matrix (\textit{e.g.}, $B(j,:)$ where $:$ is the slice operation), resulting in a intermediate row of sparse partial products (\textit{e.g.}, $C_{j,temp}(i,:)$). The addition of the sparse partial products from the multiplication of all nonzero elements in a row (\textit{e.g.}, the $i$th row) of the first input matrix with the corresponding rows of the second input matrix results in a final row of the output matrix (\textit{e.g.}, $C(i,:)=\sum\limits_{j} C_{j,temp}(i,:)$). 

The addition of sparse partial products consists of two operations: sort and merge. Each sparse row is represented by a vector of pairs $(VAL, COL\_IND)$ representing the actual value and the column index of the corresponding nonzero elements. The rows are already sorted by $COL\_IND$. To add two sorted sparse vectors, the two vectors are first sorted into a single vector based on $COL\_IND$. Then, $VAL$s of consecutive elements are merged (\textit{i.e.}, added) with the same $COL\_IND$ into a single element.

The column-wise Gustavson's method is similar to the column-wise one but with rows and columns switched. Gustavson's method does not require the hardware-expensive index-matching operation among the elements of input matrices. Also, in Gustavson's method, the addition operation of sparse partial products has low on-chip memory requirement and synchronization overhead. In contrast, since an entire intermediate matrix is dealt with using off-chip memory in the outer product method, utilizing Gustavson's method results in both improved performance and lower energy consumption. Therefore, we choose the row-wise Gustavson's method to develop FSpGEMM upon. 

However, Gustavson's method maintains sub-par reuse of the second input matrix's data due to the irregular access pattern of rows that are read from off-chip memory. Such an access pattern is dependent on the order of non-zero column indices in the rows of the first input matrix, making the caching policy or buffering scheme for the second input matrix too complex and costly to implement.

\subsection{Accelerating Dense GEMM on FPGAs}
Parametrized FPGA implementations of dot-product and matrix-vector multiplication kernels
are presented in \cite{kestur2010blas}. The work in \cite{kestur2010blas} also compared the proposed kernels with CPU and GPU implementations in terms of performance and energy efficiency.
FBLAS \cite{de2020fblas} proposes scalable, modular, and OpenCL-based implementations of the Basic Linear Algebra Subprograms (BLAS) library to improve the reusability of the FPGA kernels. This work on accelerating the BLAS library targets dense vectors and matrices thus are not suitable nor efficient for accelerating SpGEMM.

\subsubsection{Accelerating SpGEMM on FPGAs}
The work in \cite{lin2013design} and \cite{jamro2014algorithms} proposes the design and implementation of the inner product algorithm for SpGEMM. They study the performance and energy consumption trade-off of the design by tuning the architectural parameters (\textit{i.e.}, the number of PEs and the block size in blocking decomposition). Changing architectural parameters results in different FPGA resource utilization. All the existing work adopts the inner product method for SpGEMM, thus suffers from undesired performance and energy consumption overheads due to the costly index matching operation. Differently, we adopt Gustavson's algorithm eliminating expensive index matching operation, zero output computations, and storage of large intermediate values resulting in significant performance and consequently energy efficiency improvements.

\subsection{Accelerating SpGEMM on Application-specific Integrated Circuits (ASICs)}
There has been some recent work on accelerating SpGEMM on ASICs using either the outer product or Gustavson's method. OuterSPACE \cite{pal2018outerspace,park20207} and SpArch \cite{zhang2020sparch} utilizes the outer product method. SpArch reduces the number of partial output matrices by matrix condensing to mitigate the overheads of synchronization and off-chip memory access. However, both OuterSPACE and SpArch still require off-chip memory access to store and retrieve intermediate results (partial output matrices). MatRaptor \cite{srivastava2020matraptor} uses the Gustavson's method to accelerate SpGEMM. Nonetheless, this work suffers from poor data reuse of the input matrices, leading to unnecessary off-chip memory access for loading input matrices repetitively with a large performance penalty. Moreover, the scalar-vector multiplication and the addition of sparse vectors in MatRaptor \cite{srivastava2020matraptor} are performed with very low computational parallelism (1 DSP slice per PE), leading to limited performance.

\subsection{CPU and GPU Implementations of GEMM and SpGEMM}
Intel Math Kernel Library (MKL) \cite{wang2014intel} is highly optimized for Intel processors and includes threaded kernels, such as GEMM and SpGEMM, suitable for use with computationally intensive applications. MKL uses AVX, AVX2, AVX-512 SIMD instructions of Intel processors to improve performance. CSparse \cite{davis2006direct} is a sparse matrix package included in SuiteSparse. SuiteSparse is a suite of sparse matrix algorithms\cite{suite} that consists of sparse matrix operations. However, its SpGEMM kernel is single-threaded and does not utilize SIMD instructions.

The cuSPARSE library \cite{naumov2010cusparse} contains a set of basic linear algebra subroutines used for handling sparse matrices implemented on top of the NVIDIA CUDA runtime as part of the CUDA Toolkit and is designed to be invokable from C and C++. MKL and cuSPARSE contain the CPU and GPU implementations of SpGEMM, respectively, and are used as the reference SpGEMM computing solutions (in the context of HPC) for comparison in this study.

\section{Compressed Sparse Vector (CSV) Format}\label{sec:idea}

\begin{figure}[t!]
\centering
\includegraphics[width=\linewidth]{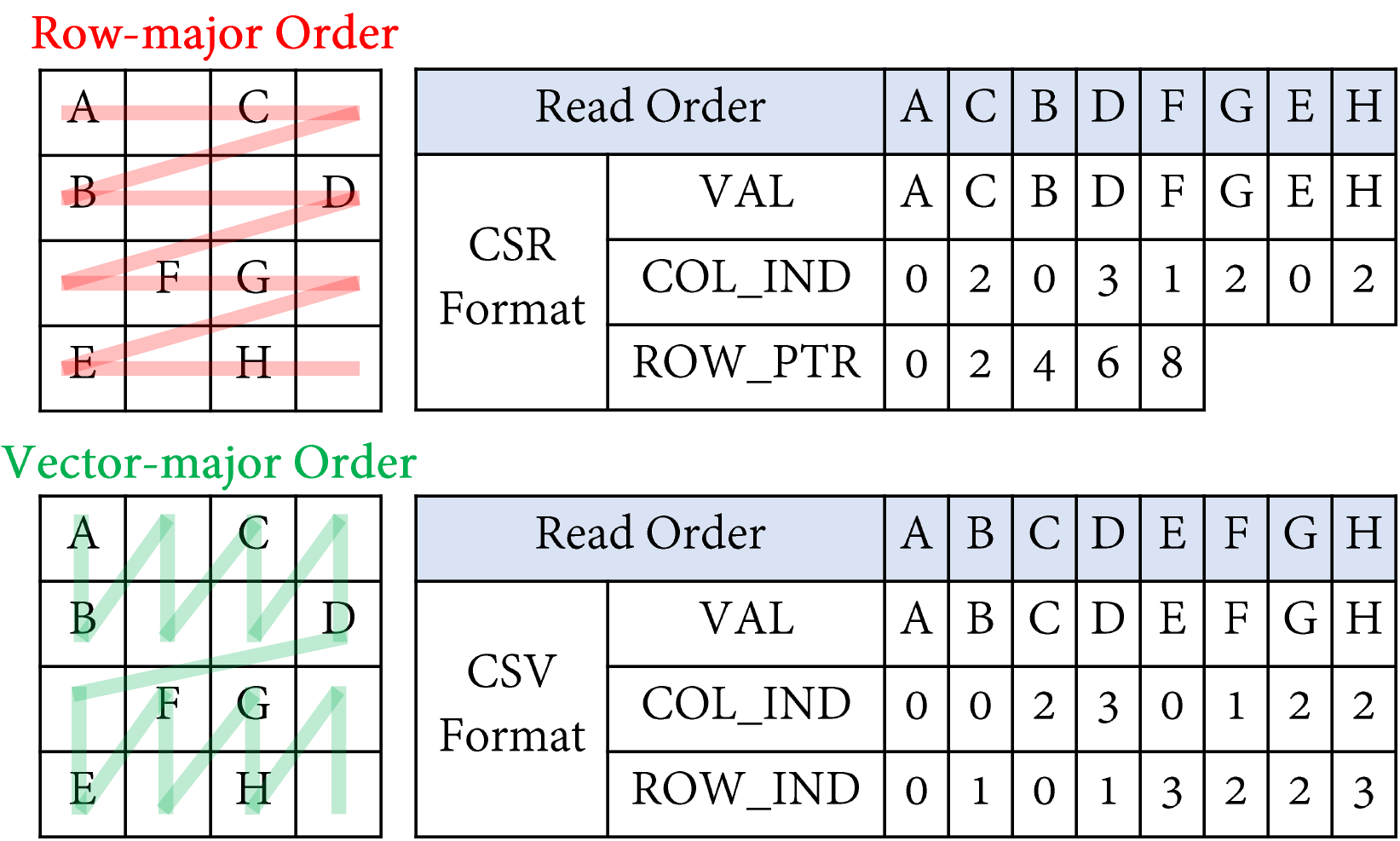}
\caption{The sparse matrix representation using the CSR and CSV formats. Each CSV vector is in the length of the number of computing units (2 assumed in this example). Transparent lines show the storage order for each format.}
\label{fig:format}
\end{figure}
As GEMM and SpGEMM are primarily computed with spatial parallelism in CPU, GPU, and most custom hardware accelerators, the input matrices are often read from off-chip memory and fed into the computation units one vector at a time, where the sparse vector length is equal to the number of the computation units. In the row-wise Gustavson's method, reading the nonzero elements from multiple rows of the input matrix based on the CSR format in a vector fashion (see Figure \ref{fig:format}) leads to a non-continuous off-chip memory access pattern. The same problem exists for the column-wise Gustavson's method when reading multiple columns of the input matrix using based on the CSC format \cite{bulucc2009parallel}. A non-continuous off-chip memory access pattern results in a large performance penalty. Additionally, using the CSC and CSR formats for the row- and column-wise Gustavson's algorithms, respectively, requires having a very large lookup table (in the size of the input matrix dimension) to keep track of the locations of the last nonzero entry read from the input matrix, which is too costly to implement for large matrices. To address these issues, we propose the CSV format tailored to Gustavson's algorithm for reading input matrices in a vector fashion. Such a vector in the length of the number of computation units is referred to as a CSV vector in the context of the CSV format.  

The CSV format uses three attributes for representing each nonzero element: the value $VAL$, the row index $ROW\_INDEX$, and the column index $COL\_INDEX$. Therefore, the location of the last nonzero entry read from the input matrix is always clearly indicated as ($ROW\_INDEX, COL\_INDEX$) without needing a loop-up table for storage. In addition, the nonzero elements in the CSV format are stored in a new vector-major order (see Figure \ref{fig:format}), which is a key difference from the CSR or CSC format that adopts a row- or column-major order. Storing nonzero elements in a vector-major order (that matches the number of computational units) is the key to assuring a continuous off-chip memory access pattern, which improves both the off-chip memory bandwidth utilization and computational performance.

\section{Framework Design}\label{sec:design}
The FSpGEMM framework consists of an FPGA kernel written in OpenCL and a host program running on an FPGA accelerator and a host CPU, respectively. FSpGEMM utilizes a new buffering scheme tailored to Gustavson's algorithm to reduce the amount of off-chip memory access. The kernel code implemented on the FPGA accelerator performs the SpGEMM computation based on Gustavson's algorithm. On the host side, we provide the utility functions for pre-processing and storing raw matrix files into the CSV format and the OpenCL API for provisioning tasks running on the FPGA accelerator.

\subsection{Data Buffering Scheme}
\begin{figure}[t!]
\centering
\includegraphics[width=0.65\linewidth]{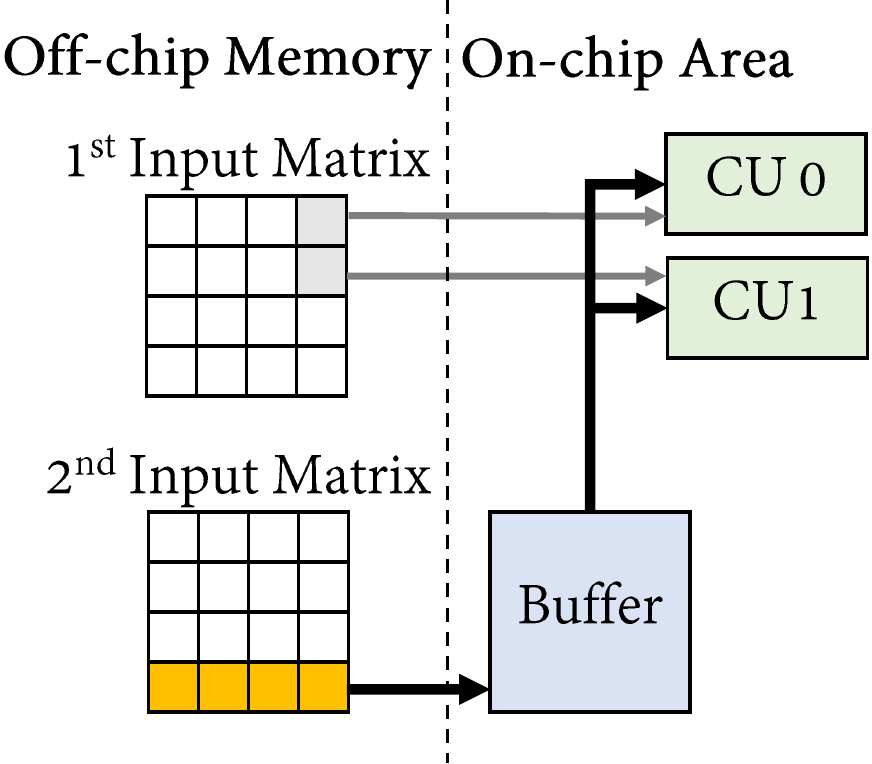}
\caption{An example for the proposed data buffering scheme. Color elements show the corresponding CSV vector and row of the first and second input matrix being processed. Each CSV vector is in the length of the number of computing units (2 CUs assumed in this example).}
\label{fig:omar}
\end{figure}

We adopt the row-wise Gustavson's method as the baseline algorithm. To improve the data reuse of the second input matrix and reduce the amount of off-chip memory access, we propose to process multiple rows of the first input matrix in parallel using multiple computing units (\textit{i.e.}, one row per computing unit) while sharing a row of the second input matrix among all computing units. To enable that, a CSV vector of nonzero values of the first input matrix is read. Based on the column index of the vector, the corresponding row from the second input matrix is then read and buffered in the on-chip memory. As a result, the access to the buffered row of the second input matrix is reused for multiple rows of the first input matrix.

The example in Figure \ref{fig:omar} shows how the proposed data buffering scheme avoids one round of off-chip memory access to $4th$ row (since the grayed out CSV vector is in the $4th$ column of the first input matrix) of the second input matrix by reusing this buffered row for all computing units.

Since for each nonzero element of the first input matrix, a row of the second input matrix is read and reused in the burst mode, we define the total number of off-chip memory access to the second input matrix as the total number of nonzero elements in the first input matrix. Consequently, We define off-chip memory access reduction (OMAR) percentage as
\begin{equation} 
OMAR (\%) = \frac{\sum\limits_{v \in V} (nnz(A(v))-1)}{nnz(A)} \times 100
\label{eq:mar}
\end{equation}

where $v$ is the CSV vector index, $V$ is a set of indices of all nonzero CSV vectors (shown as colored in Figure \ref{fig:format}), $A(v)$ is a nonzero CSV vector indexed by $v$, and $nnz(\cdot)$ is the operator that measures the number of nonzero elements in a sparse vector or matrix.

Note that the proposed data buffering scheme can be implemented on all customizable hardware (\textit{e.g.}, FPGA and ASIC devices). Also, it can potentially be applied to any fixed-architecture device with a compatible memory hierarchy.

\subsection{Hardware Architecture of the FPGA Kernel}
\subsubsection{Architectural Overview}
\begin{figure}[t!]
\centering
\includegraphics[width=\linewidth]{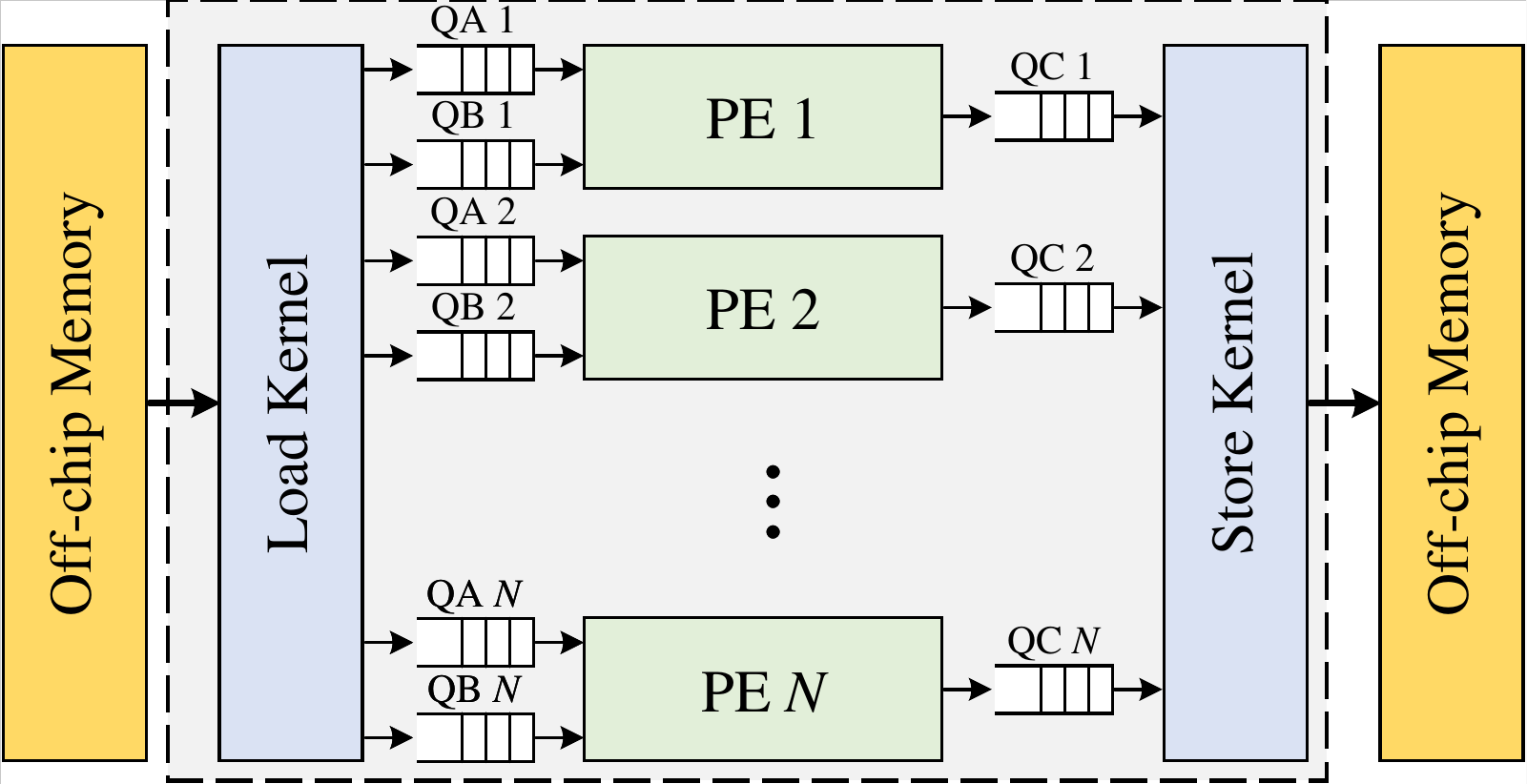}
\caption{The high-level block diagram of the hardware architecture.}
\label{fig:arch}
\end{figure}

Figure \ref{fig:arch} shows a high-level block diagram of the hardware architecture of the FPGA kernel, including three types of kernels: PEs, load and store kernels. The PEs are responsible for computing the nonzero elements of the output matrix based on Gustavson's algorithm. The load and store kernels read and write input and output data from and to the off-chip memory as well as feed and receive data to and from the PEs via FIFO channels, respectively. Separating the load and store kernels from PEs and connecting them using FIFO channels facilitates the data flow control for balancing the off-chip memory bandwidth and the data processing throughput by PEs. The depth of the FIFO channels is optimized by the offline compiler such that the load and store kernels are able to continuously read and write data from and to the off-chip memory, respectively, while pushing data throughout the kernel pipeline. 

Each PE is responsible for calculating one row of the output matrix at a time. Multiple PEs process multiple rows independently in parallel. Each PE receives a nonzero scalar value of the first input matrix along with additional scheduling information through a $QA$ channel. Also, each PE gets nonzero values from the corresponding row of the second input matrix from a $QB$ channel. The output results are streamed to the store kernel via a $QC$ channel.

To parallelize the computation of multiplications in a vector fashion, we take advantage of the loop unrolling optimization directive provided by OpenCL Offline Compiler. The FPGA kernel in FSpGEMM includes two types of OpenCL kernels: 1) \textit{autorun} kernel, which starts executing automatically and does not need to be managed by a host program, and 2) host-driven kernel. As the offline compiler provided by Intel FPGA SDK for OpenCL does not need to implement any glue logic for communicating with the host program, using \textit{autorun} kernels reduces logic utilization and allows for the mapping of more computation resources for improved performance\cite{intelpg}. The PEs are \textit{autorun} kernels, and the load and store kernels are host-driven kernels since they access the off-chip memory that must be allocated first by a host program. Overall, the FPGA kernel is parameterized with two architectural parameters: $SW$ (SIMD width of PEs) and $NUM\_PE$ (number of PEs), which allows users to balance the trade-off between computational parallelism and FPGA resource usage. 

\subsubsection{Load and Store Kernels}
One should note that the first input matrix and the output matrix are represented by the CSV format, and the second input matrix is represented by the CSR format in FSpGEMM. The load kernel iterates over the total number of nonzero elements in the first input matrix. In each iteration, the load kernel first sends a data structure ($A\_DS$) containing the value of a nonzero element of the first input matrix along with the other attribute data described in Table \ref{tab:job} using a $QA$ channel. Specifically, $RESET$ indicates whether the received nonzero element is the last one in the current row of the first input matrix. If so, $RESET=1$ signals the PEs to finish the computations and reset to their initial states. 

The load kernel takes advantage of the CSV format and identifies the nonzero elements of the first input matrix in the same CSV vector by comparing the column indices of two consecutive nonzero elements. If the column index of the next nonzero element is different from the column index of the current nonzero element, the load kernel buffers the nonzero element of the corresponding (based on the current nonzero element current index) row of the second matrix and sends it to the corresponding PEs based on the row indices of nonzero elements of the first input matrix using the CSV format. The load kernel also calculates the number of nonzero elements in a row of the second input matrix that the PEs need to expect from the load kernel. Since the values from a row of the second input matrix are loaded in a vector fashion and get fed into all the PEs, the load kernel sends $NUM\_B\_VEC$ vectors of nonzero values each in the size of $SW$ in parallel. Additionally, the row index of the nonzero element from the first input matrix ($A\_ROW\_IND$) is sent to the PEs to determine the row index of the output matrix.

Since the load kernel reads a row of the second input matrix at a time, the data are read from the off-chip memory using the CSR format to enable contiguous and regular access. Each vector of the second input matrix is represented by the data structure $B\_DS$ as described in Table \ref{tab:job1}. $VAL$ and $B\_COL\_IND$ are both a vector in the size of $SW$ that contains the values and the column indices of the nonzero elements in a row of the second input matrix.

\begin{table}[t!]
\caption{Descriptions of data structure $A\_DS$ being sent from the load kernel to PEs via a $QA$ channel.}
\label{tab:job}
\begin{tabular}{|p{0.21\linewidth}|p{0.58\linewidth}|p{0.06\linewidth}|}
\hline 
Field       & Description                                                    & Type             \\ \hline \hline
$VAL$   & a nonzero value in the first input matrix                                    & float            \\ \hline
$B\_NUM\_VEC$ & the number of nonzero vectors in the corresponding row of the second input matrix & uint \\ \hline
$A\_ROW\_IND$ & the row index of the nonzero element in the first input matrix & uint \\ \hline
$RESET$      & indicator of the end of a row of the first input matrix                                       & bool         \\ \hline
\end{tabular}
\end{table}

\begin{table}[t!]
\caption{Descriptions of data structure $B\_DS$ being read by the store kernel from the PEs via a $QB$ channel.}
\label{tab:job1}
\begin{tabular}{|p{0.19\linewidth}|p{0.52\linewidth}|p{0.14\linewidth}|}
\hline 
Field       & Description                                                    & Type             \\ \hline \hline
$VAL$       & a vector of the nonzero values in the second input matrix                                       & float[SW]          \\ \hline
$B\_COL\_IND$ & a vector of the column indices of the nonzero elements in the second input matrix & uint[SW] \\ \hline
\end{tabular}
\end{table}

\begin{table}[t!]
\caption{Descriptions of data structure $C\_DS$ being read by the store kernel from the PEs via a $QC$ channel.}
\label{tab:job2}
\begin{tabular}{|p{0.2\linewidth}|p{0.59\linewidth}|p{0.06\linewidth}|}
\hline 
Field       & Description                                                    & Type             \\ \hline \hline
$VAL$       & a nonzero value of matrix $C$                                       & float          \\ \hline
$C\_ROW\_IND$ & the row index of a nonzero element in the output matrix & uint \\ \hline
$C\_COL\_IND$ & the column index of a nonzero element in the output matrix & uint \\ \hline
\end{tabular}
\end{table}

The store kernel reads the computation results of the output matrix from the PEs, which is represented by the data structure $C\_DS$, including the nonzero value and its row and column indices as described in Table \ref{tab:job2}. Since the valid output values of different PEs are not necessarily produced at the same time, the FIFO channels $QC$ are used to facilitate the data flow control and assure data integrity.

\subsubsection{Processing Elements (PEs)}
Figure \ref{fig:arch} shows the high-level block diagram of a PE. Each PE is responsible for two major operations for producing each row of the output matrix: 1) scalar-vector multiplication of a nonzero value of the first input matrix and a row of the second input matrix resulting in an intermediate row of sparse partial products and, 2) addition of sparse partial products. These functionalities are implemented by three main submodules: a vectorized multiplication (VecMult) unit for computing sparse partial products with a SIMD width of $SW$, a sort and merge (SM) unit for sorting and accumulating partial products, and a memory unit implementing a double buffering scheme for caching intermediate results. 

\begin{figure}[t!]
\centering
\includegraphics[width=\linewidth]{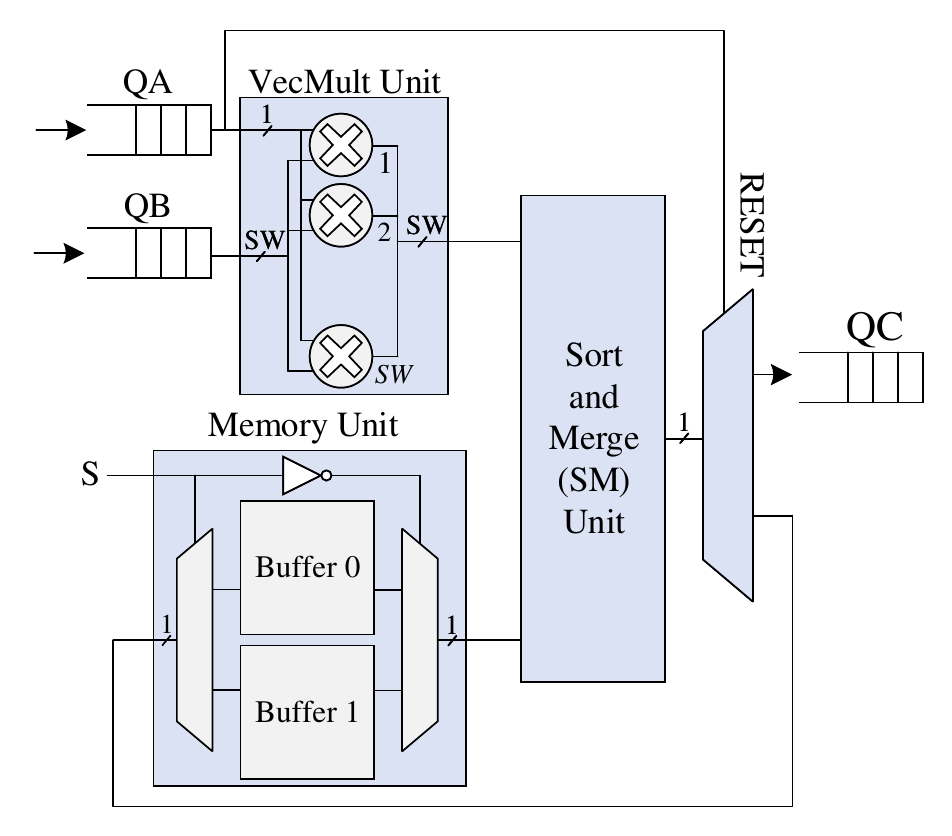}
\caption{The high-level block diagram of a PE.}
\label{fig:pe}
\end{figure}

\begin{algorithm}[t!]
\KwIn{Vectors $C\_TEMP\_VEC$ and $B\_VEC\_IND$ with size $SW$ from the VecMul unit, a $(VAL,COL\_IND)$ pair from Memory unit.}
\KwOut{A data structure $C\_DS$.} 
initialize $VEC\_PTR$ to zero for pointing to the current element of the vector\;
initialize $S$ to zero for selecting the corresponding sets of buffer from Memory unit\;
\tcc{Check if the vector is processed completely}
\While{$VEC\_PTR < SW$}{
\tcc{Check if the memory is read completely}
\eIf{$MEM\_PTR\_HEAD[S] < MEM\_PTR\_TAIL[S]$}{
        \uIf{$COL\_IND < B\_VEC\_IND[VEC\_PTR]$}{
        $C\_DS.VA=VAL$\;
        $MEM\_PTR\_HEAD[S]=MEM\_PTR\_HEAD[S]+1$\;
        }
        \uElseIf{$COL\_IND == B\_VEC\_IND[VEC\_PTR]$}{
        $C\_DS.VAL = VAL + C\_TEMP\_VEC[VEC\_PTR]$\tcc*[r]{add (merge) two elements}
        $VEC\_PTR=VEC\_PTR+1$\;
        }
        \Else{
        $C\_DS.VAL=C\_TEMP\_VEC[VEC\_PTR]$\;
        $VEC\_PTR=VEC\_PTR+1$\;
        }
    }{
        \tcc{No comparison is needed}
        $C\_DS.VAL=C\_TEMP\_VEC[VEC\_PTR]$\;
        $VEC\_PTR=VEC\_PTR+1$\;
    }
    $C\_DS.ROW\_IND=A.ROW\_IND$\;
}
\caption{The OpenCL pseudo-code implementation of the SM unit.}\label{alg:sch}
\end{algorithm}

The VecMult unit starts by reading a nonzero scalar (in the form of $A\_DS$) from a $QA$ channel and a vector (of size $SW$) of nonzero values (in the form of $A\_DS$) from a $QB$ channel for $A\_DS.B\_NUM\_VEC$ times. The VecMult unit performs $SW$ number of multiplications in parallel between $B\_VEC.VAL$ and $A\_DS.VAL$ and outputs a partial product vector $C\_TEMP\_VEC$. To reduce the on-chip memory requirement and processing latency, vector $C\_TEMP\_VEC$ is accumulated with the buffered intermediate sum of partial products $C\_TEMP\_ROW$.

To facilitate the fully pipelined addition of $C\_TEMP\_VEC$ with $C\_TEMP\_ROW$, we utilize the double-buffering technique. At any time, while $C\_TEMP\_ROW$ is being read from one of the on-chip buffers and added with $C\_TEMP\_VEC$ in the SM unit, the addition results are being stored into the other buffer. Since a sparse row of the output matrix is represented in the form of $(VAL, COL\_IND)$, two sets of double buffers are used to store the nonzero values $VAL$ and the column indices $COL\_IND$, respectively. Also, two sets of registers are used to keep track of the pointers to the head and tail of the double buffers: $MEM\_PTR\_HEAD[0]$, $MEM\_PTR\_HEAD[1]$, $MEM\_PTR\_TAIL[0]$, and $MEM\_PTR\_HEAD[1]$. The data structure $C\_TEMP\_DS$ of type $C\_DS$ is used to store the temporary computed output elements.

Algorithm \ref{alg:sch} describes the OpenCL pseudo-code implementation of the SM unit. The corresponding indices for $C\_TEMP\_VEC$ are available in $B\_VEC\_IND$. The SM unit iterates over all the elements of $C\_TEMP\_VEC$. In each iteration, it first checks if there is a valid value in the buffers to add $C\_TEMP\_VEC$ with $C\_TEMP\_ROW$ by comparing the head pointer (\textit{i.e.}, $MEM\_PTR\_HEAD[S]$) and the tail pointer (\textit{i.e.}, $MEM\_PTR\_TAIL[S]$). In this case, the SM unit compares the index of the current element of $C\_TEMP\_VEC$ with the index of stored values indicate by $MEM\_PTR\_HEAD[S]$. Since the $C\_TEMP\_VEC$ and $C\_TEMP\_ROW$ are already sorted, the value with the smaller index has the smallest index. When the smaller index exists in the buffer, the corresponding head pointer will be increased by one. In the case that the indices from $C\_TEMP\_VEC$ and $C\_TEMP\_ROW$ are equal, the merge operation occurs where the corresponding values from $C\_TEMP\_VEC$ and $C\_TEMP\_ROW$ are added. In the consequent iterations, if the head pointer is equal to the tail pointer, the SM unit has checked all the elements stored in the buffer and the values of $C\_DS$ are set based on the current $C\_TEMP\_VEC$.

All operations described in Algorithm \ref{alg:sch} happens $B\_NUM\_VEC$ times. At the end of each iteration, the double buffer selector ($S$) is switched between $0$ and $1$. Finally, there might be some remaining elements in the current buffer determined by $S$. This can be checked by comparing the head and tail pointers. All $C\_DS$ elements are set equal to the remaining values stored in the buffer with no comparison. Whenever finished processing on the vectors of the second input matrix or the remaining elements in the buffer, the PE decides whether to store $C\_DS$ in the other buffer or send the final value to a $QC$ channel based on the value of $A\_DS.RESET$.

\subsubsection{Performance Optimization for Determining Architectural Parameters}
\label{opt}
We provide a guideline to derive the optimal architectural parameters ($SW$ and $NUM\_PE$) for minimizing the runtime subject to the available resources on a given FPGA device. We formulate the runtime ($R$) as $R=\frac{N_{Ops}}{F \times P \times U}$, where $N_{Ops}$, $F$, $P$, and $U$ are the number of floating-point operations (FLOPs) to produce an output matrix, the clock frequency, the computational parallelism, and the spatial-temporal utilization factor (STUF), respectively. STUF is a statistical metric that measures on average how busy the available computation resources in a computing device are (in both space and time) for performing useful operations of a given algorithm. Thus, STUF is defined as the ratio of effective computational parallelism per clock cycle on average and ranges from 0 to 1. $N_{Ops}$ is an algorithmic parameter and depends on the dimensions and the number of nonzero elements of input matrices. Computational parallelism is determined by the total number of utilized DSP units calculated as $SW \times NUM\_PE$. Therefore, $R$ can also be expressed as
\begin{equation} 
\label{equ:run}
R=\frac{N_{Ops}}{F \times SW \times NUM\_PE \times U}=\frac{\alpha}{SW \times NUM\_PE}\textrm{,}
\end{equation}
where $N_{Ops}$, $F$, and $U$ are lumped as $\alpha$. We define the constrained optimization problem as
\begin{equation} 
\label{equ:opt}
\begin{split}
&\textrm{minimize}\quad R(SW)=\frac{\alpha}{SW \times NUM\_PE} \\
&\textrm{subject to} \quad f_1(SW) \leq C_1 \textrm{,}f_2(SW,NUM\_PE) \leq C_2 \textrm{,}
\end{split}
\end{equation}
where $f_1(SW)=sizeof(float) \times SW \times F$ and $C_1$ are the required and available off-chip memory bandwidth, respectively, with $sizeof(float)=32b$. Function $f_2(SW,NUM\_PE)=\beta \times P=\beta \times SW \times NUM\_PE$ and $C_2$ are the total required and available logic resources, respectively, where $\beta$ is a linear fitting parameter that captures the proportionality of the actual logic resource usage over computational parallelism that can be derived based on the logic resource ($f_2$) usage reported by the offline compiler given a target FPGA device. 

Based on $R(SW)=\frac{\alpha}{SW \times NUM\_PE}$, the runtime $R$ is minimized when $SW$ and $NUM\_PE$ are maximized. Thus, the constrained optimization problem defined in Equation \ref{equ:opt} has an analytical solution that can be derived as follows.
\begin{enumerate}
    \item Derive $SW$ as $SW =\lceil \frac{C_1}{sizeof(float) \times F} \rceil$.
    \item Set $NUM\_PE=1$ and run the offline compiler to derive the value of $\beta$ as $\beta = \frac{f_2}{SW \times NUM\_PE}$.
    \item Derive $NUM\_PE$ as $NUM\_PE =\lceil \frac{C_2}{\beta \times SW} \rceil$
\end{enumerate}

\subsection{Host Program}
The host program running on the CPU includes two major parts: pre-processing utility functions and OpenCL API function calls. As the first input matrix needs to be transformed and stored in the CSV format prior to the computation by the FPGA kernel, the utility functions read in the raw matrix files in an existing sparse matrix format then convert and store the matrices in the CSV format. The pre-processing step only needs to be performed once. We have utilized OpenCL API functions to create memory buffers, enqueue kernels, and read the results back from the FPGA device.

\section{Evaluation}\label{sec:eval}
\subsection{Experiment Setup}
We evaluate the performance and energy efficiency of FSpGEMM in terms of runtime ($s$) and energy consumption ($J$) per SpGEMM computation. To evaluate the FSpGEMM framework, we select a set of sparse matrices from the publicly available SuiteSparse Matrix Collection \cite{davis2011university} (formerly known as the University of Florida Sparse Matrix Collection), a collection of sparse matrices in real applications. The specifications of the matrices are summarized in Table \ref{tab:mat} including matrix dimensions (the number of rows times the number of columns) and the density calculated as $\frac{No. \ of \ Nonzero \ Elements}{Matrix \ Size}$.

\begin{table}[t!]
\caption{\label{tab:mat} The specification of matrices chosen from the SuiteSparse Matrix Collection.}
\centering
\begin{tabular}{|c|c|c|c|}
\hline
Matrix & Dimensions & Density \\ \hline
\hline
poisson3Da     & 14K $\times$ 14K & 1.9e-3 \\ \hline
2cubes\_sphere & 101K $\times$ 101K                 & 1.6e-4 \\ \hline
filter3D       & 106K $\times$ 106K                   & 2.4e-4 \\ \hline
cage12         & 130K $\times$ 130K                 & 1.2e-4 \\ \hline
scircuit       & 171K $\times$ 171K                 & 3.3e-5 \\ \hline
mac\_econ\_fwd500         & 207K $\times$ 207K                 & 3.0e-5 \\ \hline
offshore       & 260 $\times$ 260K                 & 6.3e-5 \\ \hline
webbase-1M       & 1000K $\times$ 1000K                 & 3.1e-6 \\ \hline
\end{tabular}
\end{table}

Our design adopts the single-precision floating-point (32-bit) data format. We develop and implement FSpGEMM using OpenCL with Intel FPGA SDK for OpenCL with Quartus Prime Pro 20.1. We compare FSpGEMM with MKL and cuSPARSE libraries for the CPU and GPU implementations, respectively. We measure the performance of the CPU implementation of MKL on a dual-socket Intel Xeon E5-2637 v3 CPU \cite{intelpow}, and the GPU implementation of cuSPARSE on an NVIDIA GTX TITAN X GPU. Our work is evaluated on an Intel Arria 10 GX FPGA Development Board \cite{intela10}. Table \ref{tab:spec} summarizes the specifications of the CPU system, the GPU device, and the FPGA board used in the evaluation.

\begin{table}[]
\caption{\label{tab:spec} Specifications of the CPU system, the GPU device, and the FPGA board used in the evaluation.}
\begin{tabular}{|p{0.3\linewidth}|p{0.6\linewidth}|}
\hline
Hardware Platform                             & Specification                                                                                           \\ \hline\hline
Intel Xeon E5-2637 v3 CPU                    & 15M Cache, 3.50-3.70 GHz Clock Frequency, 4 Cores, 2 Sockets, 68 GB/s Memory Bandwidth                       \\ \hline
NVIDIA GTX TITAN X GPU                       & 3,072 CUDA Cores, 1,000-1,075 MHz Clock Frequency, 336.5 GB/s Memory Bandwidth \\ \hline
Intel Arria 10 GX FPGA Board & 1,518 DSPs, 53 Mb M20K, 12.7 Mb MLAB, 15 GB/s Memory Bandwidth                                            \\ \hline
\end{tabular}
\end{table}

\subsection{OMAR Evaluation}
\begin{figure*}
     \centering
     \begin{subfigure}[b]{0.24\textwidth}
         \centering
         \includegraphics[width=\textwidth]{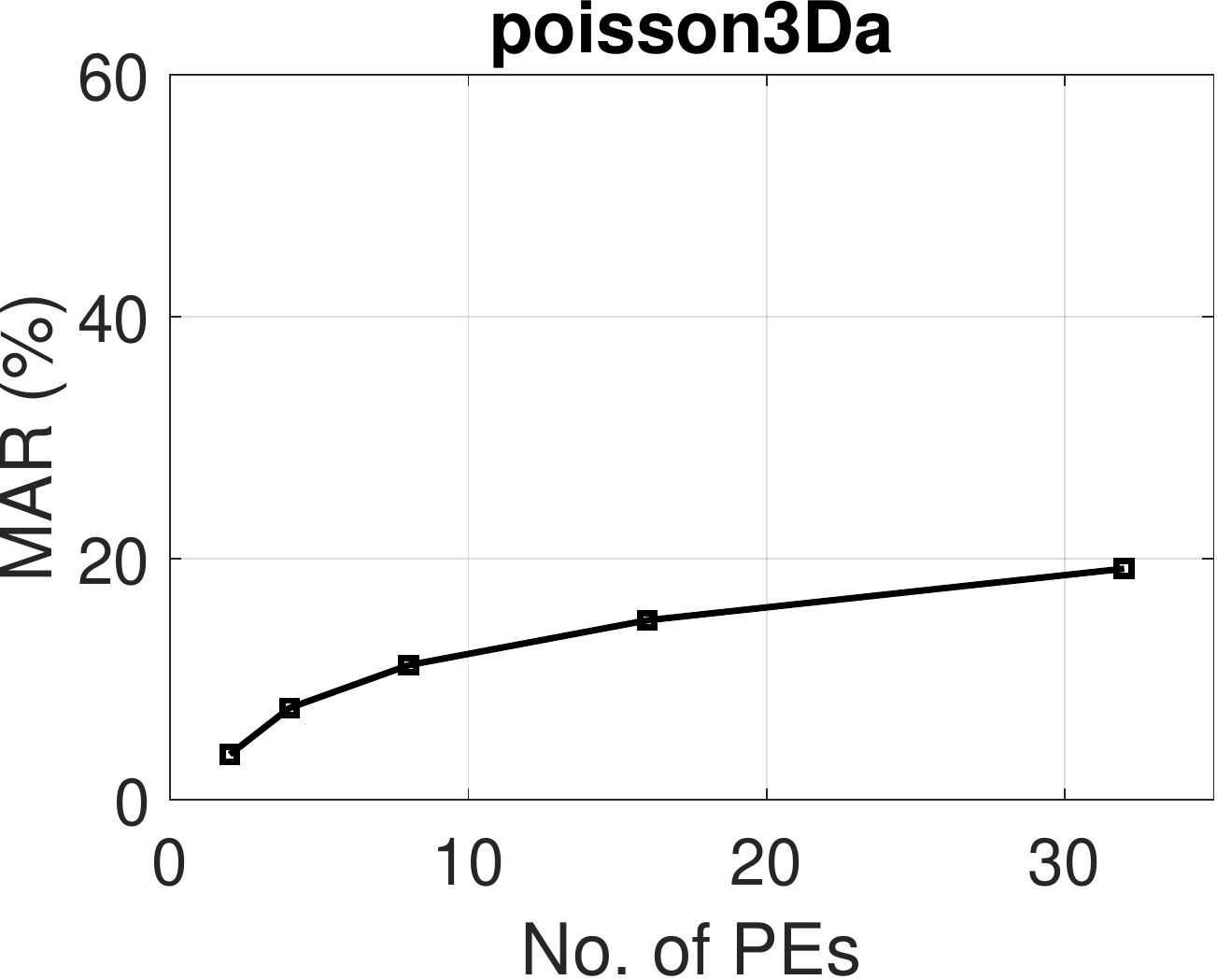}
     \end{subfigure}
     \begin{subfigure}[b]{0.24\textwidth}
         \centering
         \includegraphics[width=\textwidth]{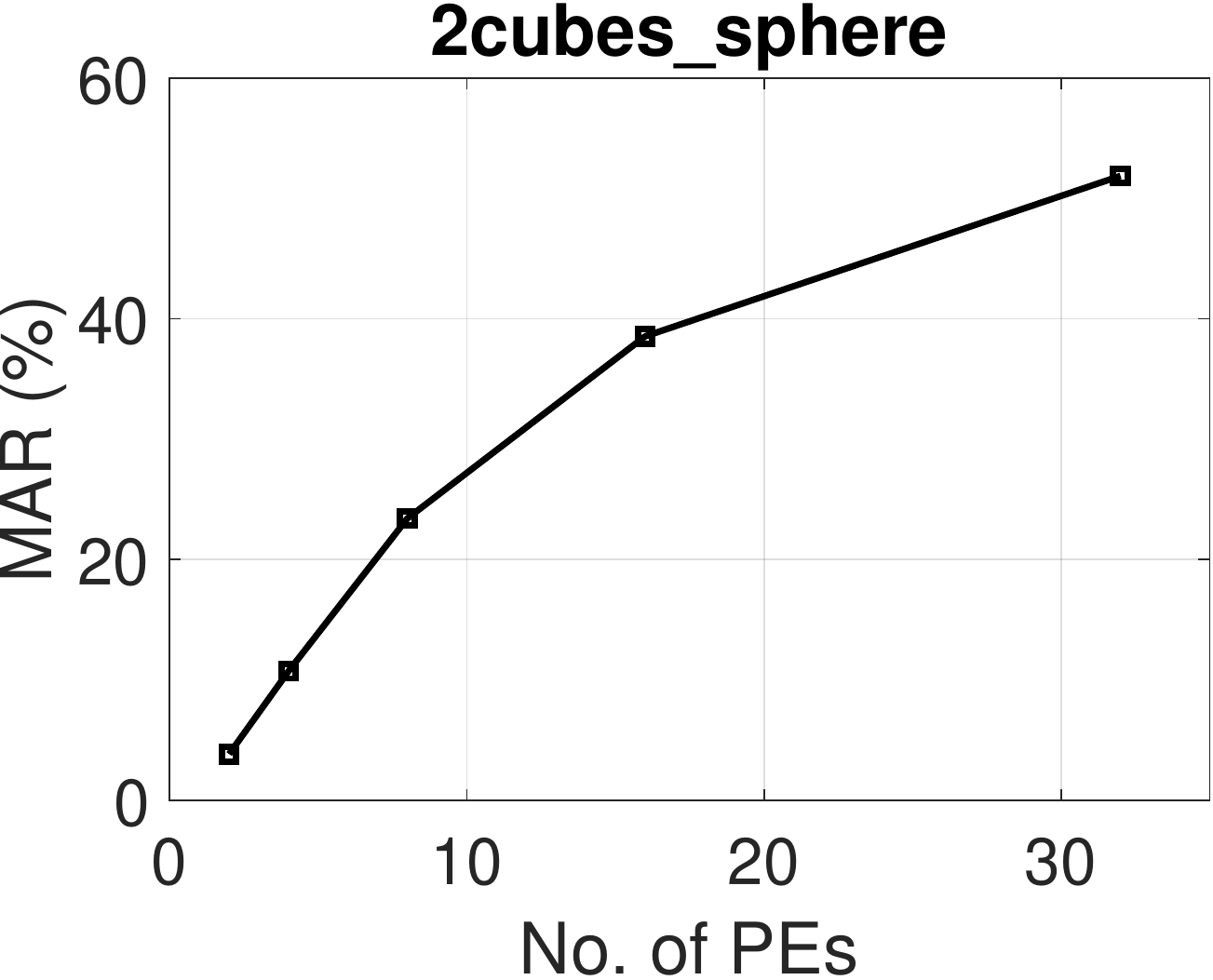}
     \end{subfigure}
     \begin{subfigure}[b]{0.24\textwidth}
         \centering
         \includegraphics[width=\textwidth]{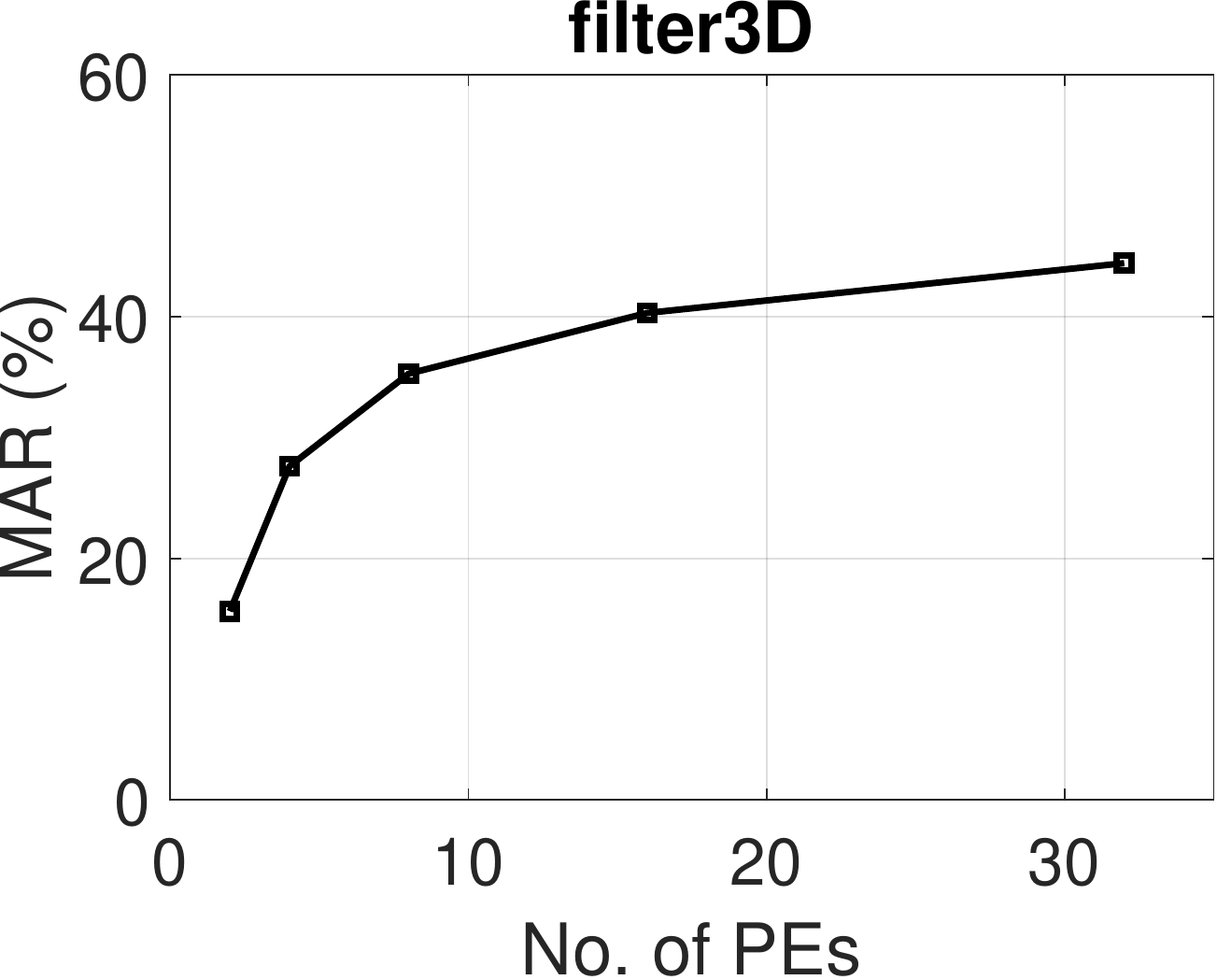}
     \end{subfigure}
     \begin{subfigure}[b]{0.24\textwidth}
         \centering
         \includegraphics[width=\textwidth]{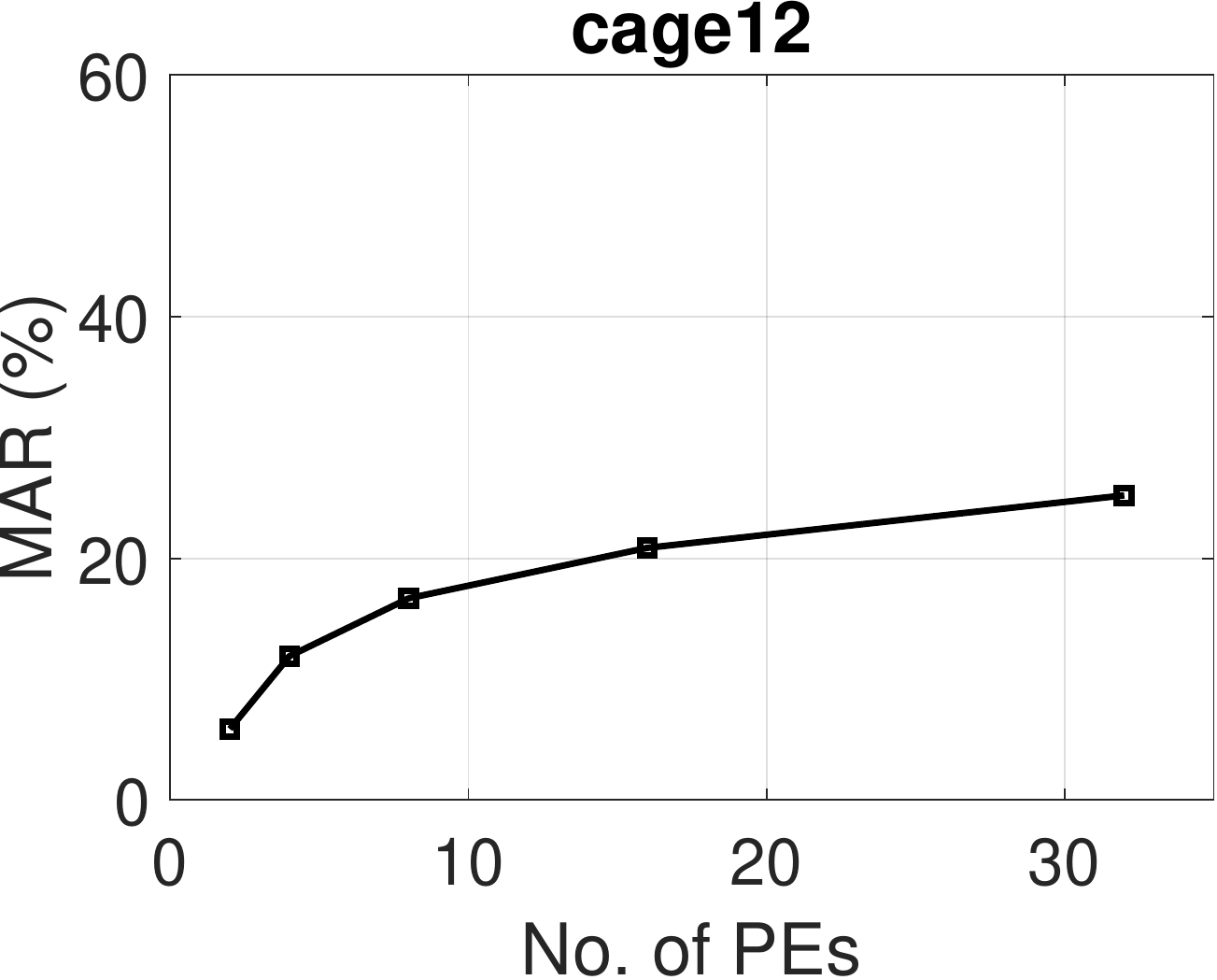}
     \end{subfigure}
     \par\bigskip
     \begin{subfigure}[b]{0.24\textwidth}
         \centering
         \includegraphics[width=\textwidth]{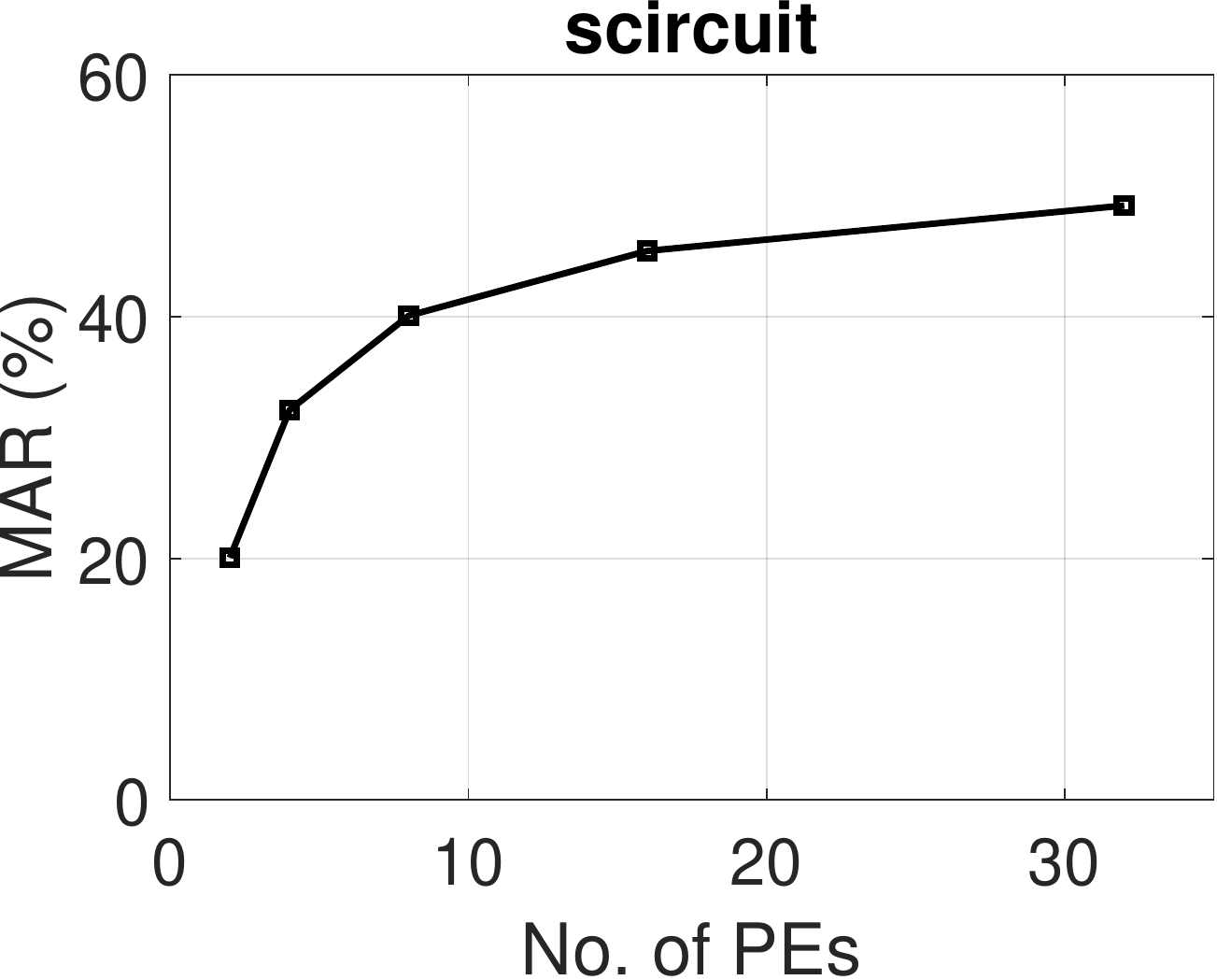}
     \end{subfigure}
     \begin{subfigure}[b]{0.24\textwidth}
         \centering
         \includegraphics[width=\textwidth]{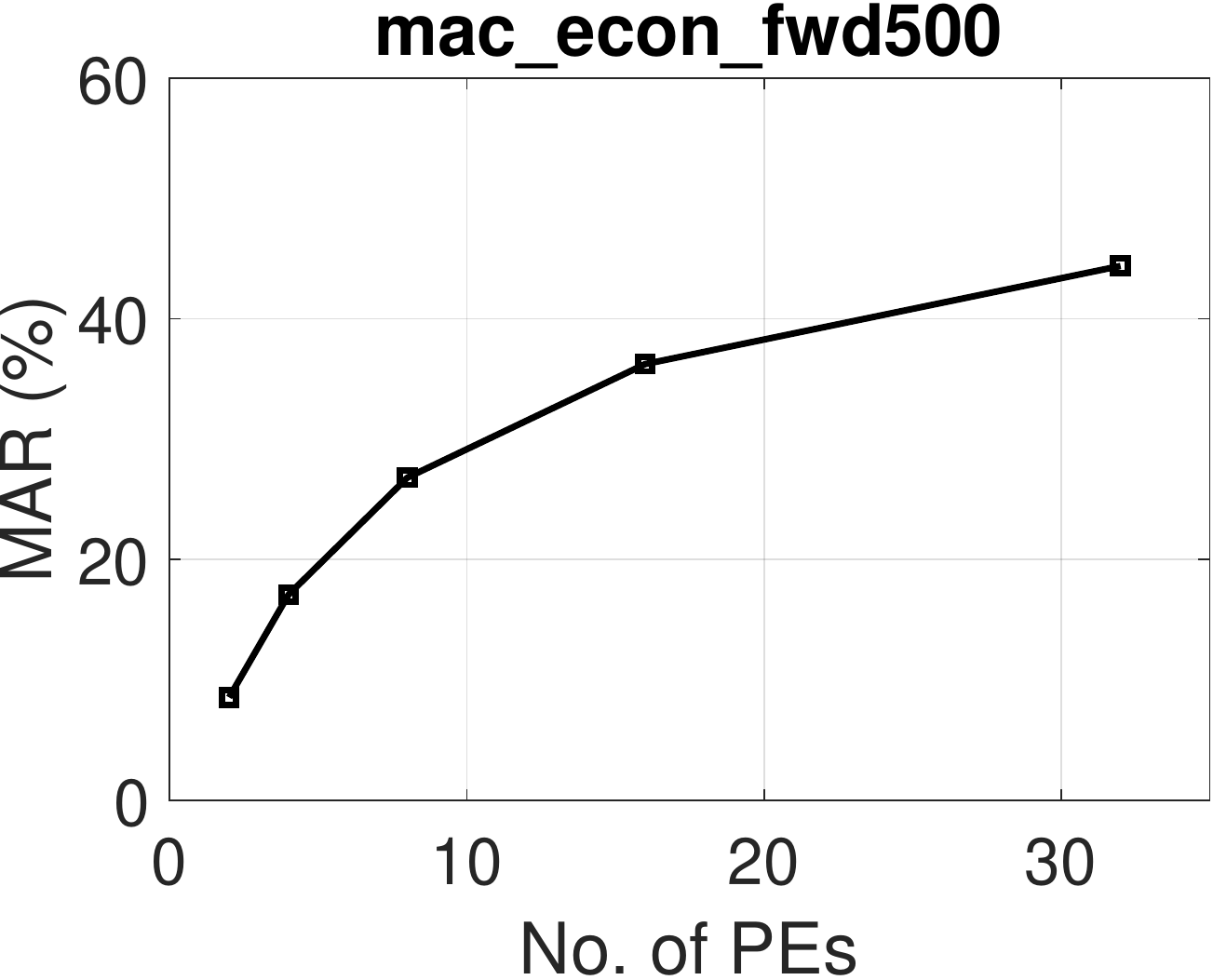}
     \end{subfigure}
     \begin{subfigure}[b]{0.24\textwidth}
         \centering
         \includegraphics[width=\textwidth]{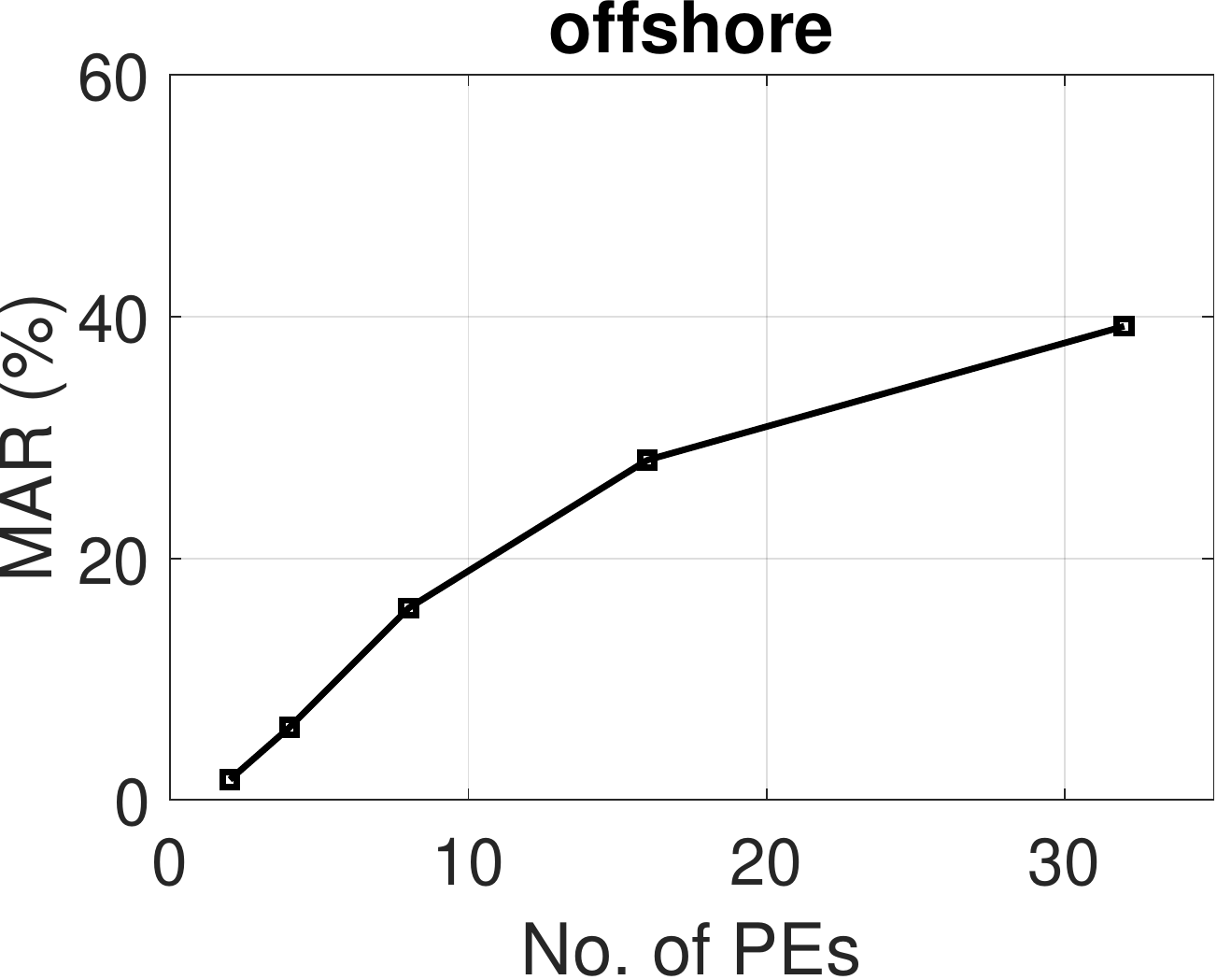}
     \end{subfigure}
     \begin{subfigure}[b]{0.24\textwidth}
         \centering
         \includegraphics[width=\textwidth]{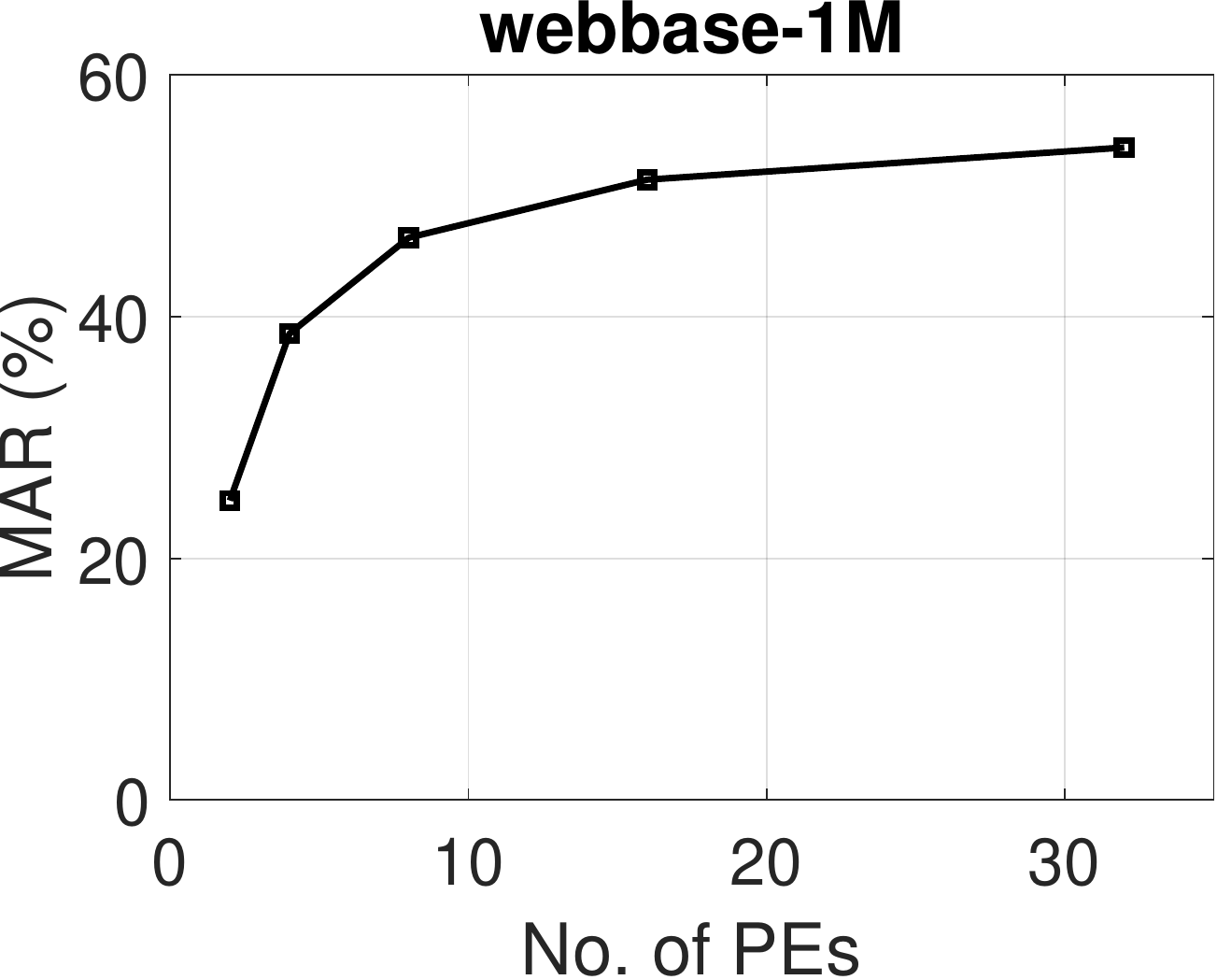}
     \end{subfigure}
        \caption{Off-chip memory access reduction (OMAR) percentage with respect to different input matrices and the number of PEs. The difference in sparsity patterns of the input matrices results in the difference in OMAR percentage.}
        \label{fig:mar}
\end{figure*}

As mentioned in Section \ref{sec:idea}, a naive implementation of Gustavson's algorithm suffers from poor reuse of the second input matrix. To address this issue, we propose to process multiple rows of the first input matrix in parallel and share the rows of the second among multiple PEs. We calculate the off-chip memory access reduction percentage (OMAR \%) using Equation \ref{eq:mar} for the matrices summarized in Table \ref{tab:mat}. Figure \ref{fig:mar} shows the OMAR percentage that can be achieved by the proposed data buffering scheme with respect to different numbers of PEs (\textit{i.e.}, parallel rows) and matrices. The results show that the amount of OMAR that can be archived by the proposed data buffering scheme monotonically improves as the number of PEs increases. 1.7\%-24.8\%, 6.0\%-38.6\%, 15.9\%-46.5\%, 28.1\%-51.3\%, and 39.2\%-54.0\% OMAR can be achieved at the PE number of 2, 4, 8, 16, and 32, respectively, across the selected sparse matrices. The improvement with increasing the number of PEs is because a row of the second input matrix is buffered and shared for processing more rows of the first input matrix. Thus, the amount of access to the row of the second input matrix is decreased.

\subsection{Experiment Results}
The optimal architectural parameters of the FPGA kernel implementation derived from the performance optimization method presented in Section \ref{opt} are $SW=16$, $NUM\_PE=32$. The FPGA kernel implementation runs at 236 MHz. Table \ref{tab:util} shows the resource utilization for the implemented FPGA kernel. As projected by our performance and architecture models, the FPGA kernel implementation fully utilizes the FPGA resources (limited by logic utilization), and further increasing $SW$ beyond $16$ does not improve performance any further due to the saturation of off-chip memory bandwidth.

\begin{table}[t!]
\centering
\caption{\label{tab:util} Resource utilization on the Intel Arria 10 GX FPGA with $SW=16$ and $NUM\_PE=32$.}
\begin{tabular}{|c|c|c|c|c|c|}
\hline
Resource & Logic & ALUTs & Registers & Memory & DSPs  \\ \hline \hline
Usage & 97\% & 60\% & 42\%  & 34\% & 36\%   \\ \hline
\end{tabular}
\end{table}

\subsubsection{Performance Comparison with CPU and GPU Implementations}
\begin{figure}[t!]
\centering
\includegraphics[width=\linewidth]{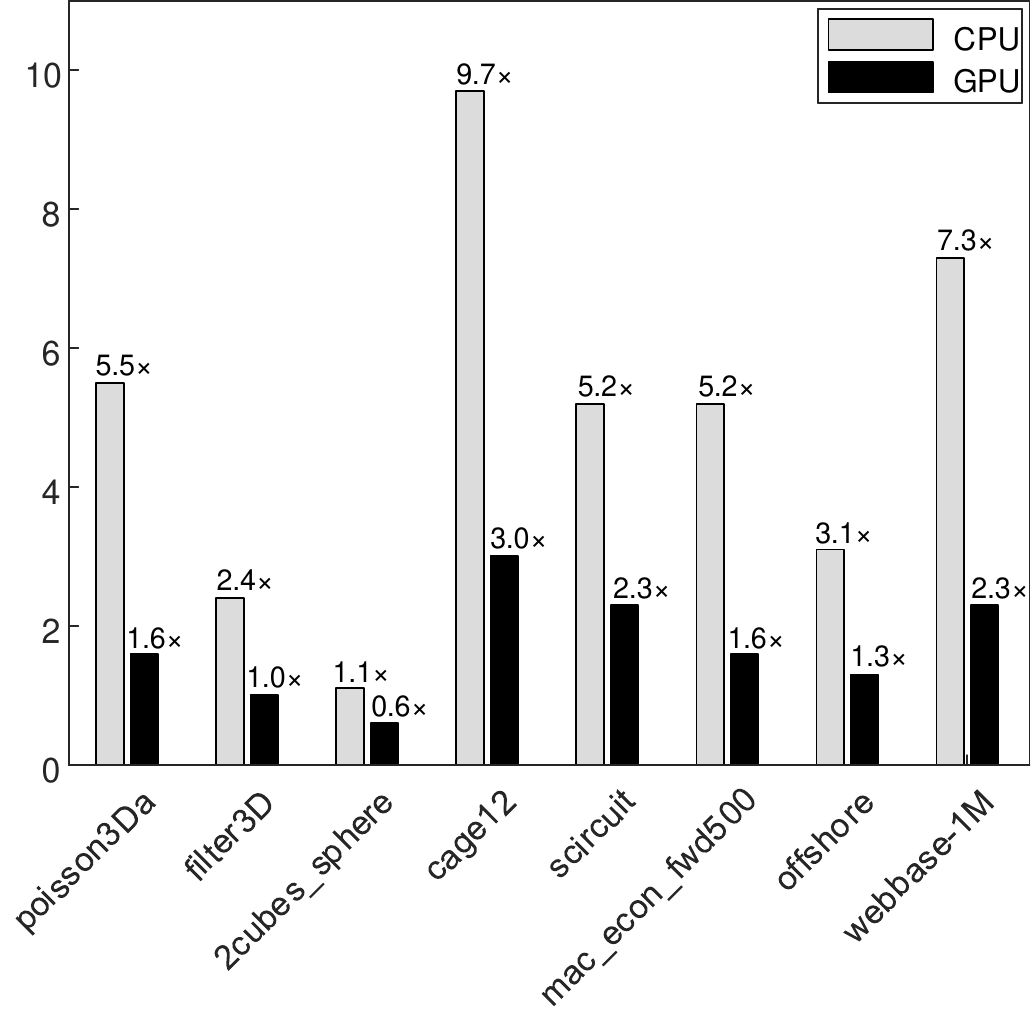}
\caption{Runtime speedup achieved by the FPGA implementation of FSpGEMM over the CPU and GPU implementations of MKL and cuSPARSE, respectively.}
\label{fig:speedup}
\end{figure}

Table \ref{tab:perf} shows the performance comparison of FSpGEMM in terms of runtime in $ms$ with the CPU implementation of MKL and the GPU implementation of cuSPARSE for performing SpGEMM computation. A lower runtime indicates higher performance. Figure \ref{fig:speedup} shows the performance speedup achieved by FSpGEMM over the CPU and GPU implementations with respect to different sparse matrices. Overall, the speedup over the CPU and GPU implementations ranges from 1.1$\times$-9.7$\times$ and 0.6$\times$-3.0$\times$, respectively. On average, FSpGEMM achieves 4.9$\times$ and 1.7$\times$ higher performance than the CPU and GPU counterparts, respectively. It should be noted that such speedup ratios are achieved at the condition that the FPGA implementation runs at a 5-15$\times$ lower clock frequency than the CPU and GPU implementations. The speedup mainly stems from the off-chip memory access reduction due to the customized data buffer scheme tailored to Gustavson's algorithm, the improved off-chip memory bandwidth utilization as a result of the architectural co-design with the proposed CSV format, the temporal/pipeline parallelism offered by FPGAs for resolving the strong data dependency in the Gustavson's algorithm, and the throughput-optimized hardware architecture tailored to the Gustavson's algorithm, all of which improves the overall STUF of the computation resources on the FPGA at run time.  

\begin{table}[t!]
\caption{\label{tab:perf} Runtime ($ms$) comparison between the FPGA implementation of FSpGEMM and the CPU and GPU implementations of MKL and cuSPARSE, respectively.}
\begin{tabular}{|p{0.26\linewidth}|p{0.19\linewidth}|p{0.19\linewidth}|p{0.19\linewidth}|}
\hline
Matrix                    & MKL (CPU)   & cuSPARSE (GPU)                     & FSpGEMM        \\ \hline \hline
poisson3Da        & 27   & 8  & 5  \\ \hline
2cubes\_sphere    & 21   & 9  & 9 \\ \hline
filter3D          & 44  & 25 & 42 \\ \hline
cage12            & 147  & 46 & 15 \\ \hline
scircuit          & 32   & 14 & 6 \\ \hline
mac\_econ\_fwd500 & 36   & 11 & 7 \\ \hline
offshore          & 71   & 30 & 23 \\ \hline
webbase-1M        & 181 & 57 & 25 \\ \hline
\end{tabular}
\end{table}

\subsubsection{STUF Comparison with CPU and GPU Implementations}
In section \ref{sec:design}, we defined the runtime as $R=\frac{N_{Ops}}{F \times P \times U}$, where $N_{Ops}$, $F$, $P$, and $U$ are the number of FLOPs in the Gustavson's algorithm, clock frequency, computational parallelism, and STUF, respectively. We consider computational parallelism as the number of possible floating-point operations per clock cycle. A DSP in the FPGA and a CUDA core in the GPU \cite{nvidia2009nvidia} can perform one multiplication and one addition per clock cycle. Therefore, the computational parallelism for the FPGA and GPU is equal to twice the number of DSPs (1,518 DSPs) and twice the number of CUDA cores (3,584 cores), respectively. The CPU has two sockets with four cores per socket, and each core provides 32 FLOPs per cycle, enabled by Intel AVX2 instruction set extension \cite{lento2014optimizing}. Therefore, we can estimate the actual STUF of different solutions based on the runtime measurements as $U=\frac{N_{Ops}}{F \times P \times R}$.

Table \ref{tab:util} compares the STUF of the FPGA implementation of FSpGEMM with the CPU implementation of MKL and the GPU implementation of cuSPARSE for running the SpGEMM algorithms, and the numbers in parentheses show the STFU improvement of FCHOL with respect to the corresponding matrix. Overall, the STUF improvement over the CPU and GPU implementations ranges from 1.4$\times$-12.4$\times$ and 5.1$\times$-26.0$\times$, respectively. On average, FSpGEMM achieves a 6.3$\times$\ and 14.7$\times$ higher STUF than CPU and GPU counterparts, respectively.

\begin{table}[t!]
\caption{\label{tab:util} STUF comparison between the FPGA implementation of FSpGEMM and the CPU and GPU implementations of MKL and cuSPARSE, respectively.}
\begin{tabular}{|p{0.26\linewidth}|p{0.19\linewidth}|p{0.19\linewidth}|p{0.15\linewidth}|}
\hline
Matrix                    & MKL   & cuSPARSE                     & FSpGEMM        \\ \hline \hline
poisson3Da        & 4.7E-04 (7.1$\times$)  & 2.4E-04 (14.1$\times$) & 3.4E-03 \\ \hline
2cubes\_sphere    & 1.4E-03 (3.0$\times$)  & 5.0E-04 (8.6$\times$) & 4.3E-03 \\ \hline
filter3D          & 2.1E-03 (1.4$\times$)   & 5.6E-04 (5.1$\times$)  & 2.9E-03 \\ \hline
cage12            & 2.6E-04 (12.4$\times$)  & 1.2E-04 (26.0$\times$) & 3.2E-03 \\ \hline
scircuit          & 2.9E-04 (6.6$\times$)  & 1.0E-04 (19.3$\times$) & 2.0E-03 \\ \hline
mac\_econ\_fwd500 & 2.3E-04 (6.7$\times$)  & 1.1E-04 (13.7$\times$) & 1.5E-03 \\ \hline
offshore          & 1.2E-04 (4.0$\times$)   & 4.1E-05 (11.2$\times$) & 4.6E-04 \\ \hline
webbase-1M        & 4.2E-04 (9.4$\times$) & 2.0E-04 (19.8$\times$) & 3.9E-03 \\ \hline
\end{tabular}
\end{table}

\begin{table}[h]
\caption{\label{tab:energy} Energy consumption ($J$) comparison between the FPGA implementation of FSpGEMM and the CPU and GPU implementations of MKL and cuSPARSE, respectively.}
\begin{tabular}{|p{0.26\linewidth}|p{0.19\linewidth}|p{0.19\linewidth}|p{0.19\linewidth}|}
\hline
Matrix                    & MKL (CPU)   & cuSPARSE (GPU)                     & FSpGEMM        \\ \hline \hline
poisson3Da        & 3.46   & 1.31 & 0.09 \\ \hline
2cubes\_sphere    & 3.11   & 1.22 & 0.17 \\ \hline
filter3D          & 6.03   & 3.43 & 0.79 \\ \hline
cage12            & 16.91   & 6.44 & 0.29 \\ \hline
scircuit          & 4.35   & 1.83 & 0.12 \\ \hline
mac\_econ\_fwd500 & 5.22   & 1.43 & 0.13 \\ \hline
offshore          & 9.80   & 3.99 & 0.44 \\ \hline
webbase-1M        & 15.93 & 9.86 & 0.47 \\ \hline
\end{tabular}
\end{table}

\subsubsection{Energy Efficiency Comparison with CPU and GPU Implementations}
We measure the average power consumption of the Intel Arria 10 GX development board using the Power Monitor tool in the Board Test System (BTS) application provided by Intel \cite{intelug} during FPGA kernel execution. BTS measures the supply voltage and the drawn current of the entire FPGA board by reads values from onboard sensors.

For the power measurement of the CPU, we utilize the likwid-powermeter tool from Likwid \cite{gruber2019likwid} to access the Running Average Power Limit (RAPL) counters on the Intel CPU. The RAPL interface is controlled through MSR registers \cite{rapl}. For the power measurement of the GPU, we utilize the POWER query option \cite{nsmidoc} of the NVIDIA System Management Interface (\textit{nvidia-smi}) \cite{nsmi} tool. 

\begin{figure}[t!]
\centering
\includegraphics[width=\linewidth]{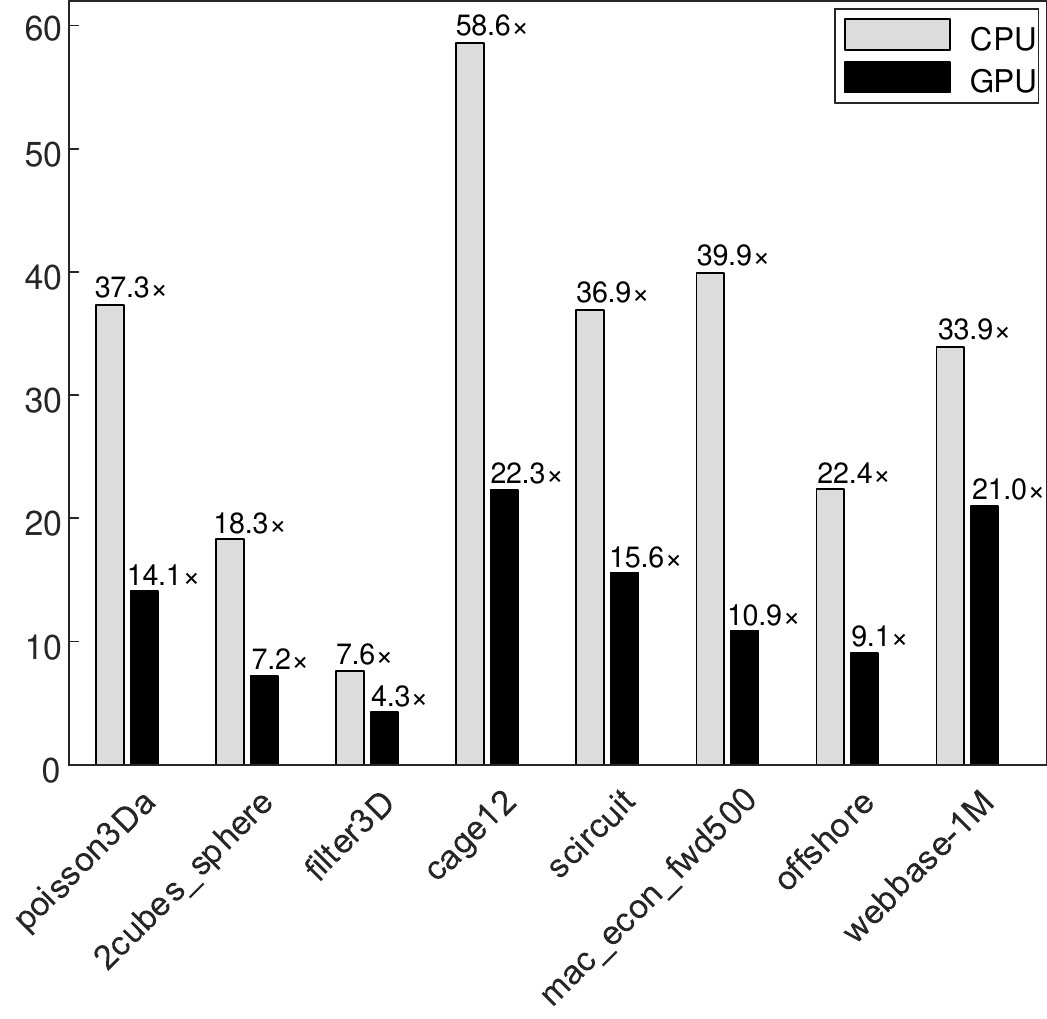}
\caption{Energy consumption reduction achieved by the FPGA implementation of FSpGEMM over the CPU and GPU implementations of MKL and cuSPARSE, respectively.}
\label{fig:reduction}
\end{figure}

Table \ref{tab:energy} summarizes the energy consumption ($J$) per SpGEMM computation of different implementations, which is calculated as product of runtime ($s$) and average power consumption ($W$). Figure \ref{fig:reduction} shows the corresponding energy consumption reduction achieved by the FPGA implementation. Overall, the energy consumption reduction over the CPU and GPU implementations ranges from 7.6$\times$-58.6$\times$ and 4.3$\times$-22.3$\times$, respectively. On average, 
FSpGEMM achieves 31.9$\times$ and 13.1$\times$ higher energy efficiency than the CPU and GPU counterparts, respectively.

\section{Conclusion}\label{sec:con}
In this paper, we present FSpGEMM, an OpenCL-based HPC framework for FPGA acceleration of SpGEMM. FSpGEMM includes a deeply pipelined and scalable FPGA kernel that accelerates the SpGEMM algorithm and a set of utility functions for pre-processing step of input matrices into the proposed CSV format. 

We introduce a performance-optimized model to derive architectural parameters for the FPGA kernel subject to the available on-chip resources (DSPs and memory blocks) to map the design into the arbitrary FPGAs. The experiment results based on the Intel Arria 10 GX FPGA development board for accelerating SpGEMM of a set of sparse matrices from SuiteSparse Matrix Collection show on average one order of magnitude higher performance and lower energy consumption compared to the state-of-the-art implementations of SpGEMM on CPU and GPU.

\bibliographystyle{IEEEtran}
\bibliography{sample-base.bib}

%

\begin{IEEEbiography}[{\includegraphics[width=1in,height=1.25in,clip,keepaspectratio]{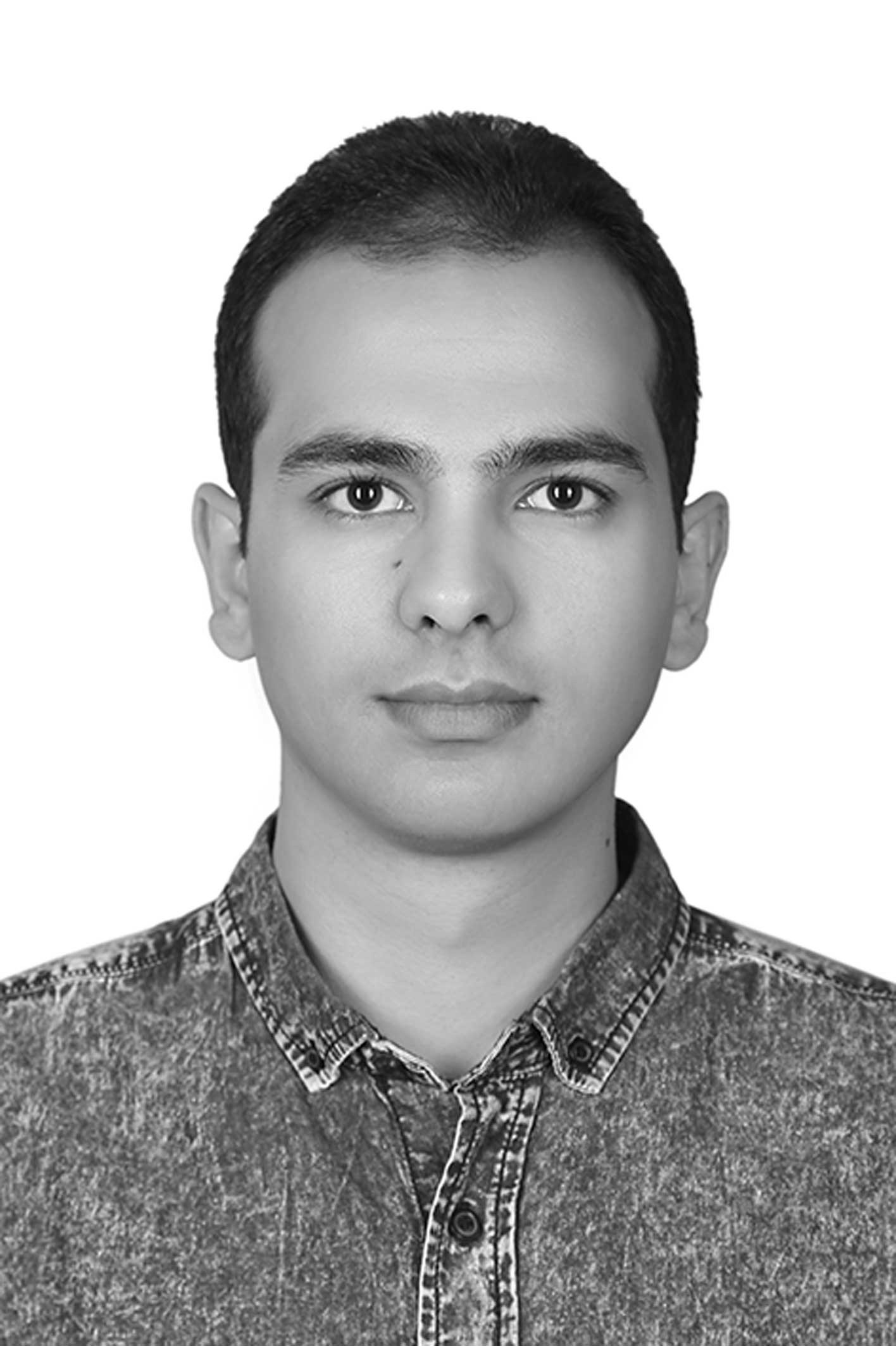}}]{Erfan Bank Tavakoli}
Erfan Bank Tavakoli received the B.Sc. degree from Iran University of Science and Technology, Tehran, Iran, in 2017, the M.Sc. degree from the University of Tehran, Tehran, Iran, in 2019, all in electrical engineering. He is currently a computer engineering Ph.D. student at Arizona State University, AZ, USA. His current research interests include high performance computing and hardware acceleration.
\end{IEEEbiography}

\begin{IEEEbiography}[{\includegraphics[width=1in,height=1.25in,clip,keepaspectratio]{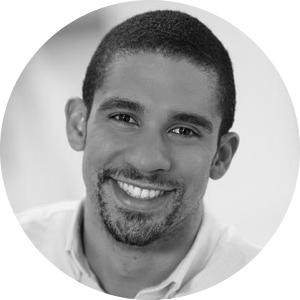}}]{Michael Franklin Riera}
Michael Franklin Riera Received his B.Sc from DeVry University in 2005, and MS degree from the University of Central Florida in 2012. He is currently a Computer Engineering PhD student at Arizona State University, USA. Michael's current research involves system software and accelerator IP for exascale heterogeneous computing.
\end{IEEEbiography}

\begin{IEEEbiography}[{\includegraphics[width=1in,height=1.25in,clip,keepaspectratio]{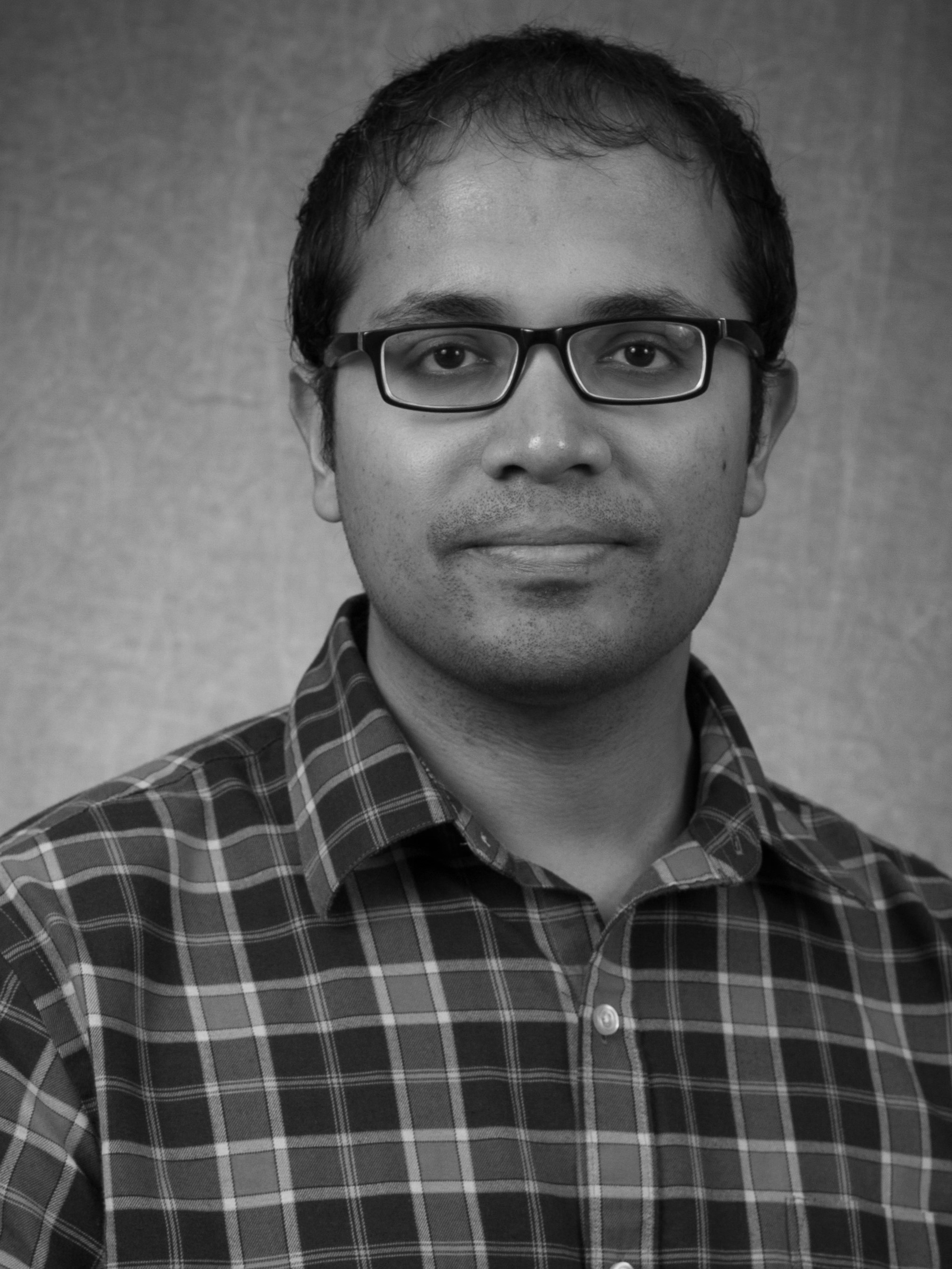}}]{Masudul Hassan Quraishi}
Masudul Hassan Quraishi received the B.Sc. Degree in Electrical Engineering from Bangladesh University of Engineering and Technology, Bangladesh in 2013 and MS degree in Computer Engineering from Arizona State University, USA in 2020. He is currently a Computer Engineering PhD student at Arizona State University, USA. His current research involves virtualization of FPGA for high performance computing. 
\end{IEEEbiography}

\begin{IEEEbiography}[{\includegraphics[width=1in,height=1.25in,clip,keepaspectratio]{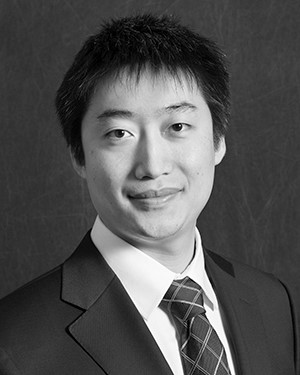}}]{Fengbo Ren}
Fengbo Ren (S’10-M’15-SM’20) received the B.Eng. Degree from Zhejiang University, Hangzhou, China, in 2008 and the M.S. and Ph.D. degrees from the University of California, Los Angeles, in 2010 and 2014, respectively, all in electrical engineering.

In 2015, he joined the faculty of the School of Computing, Informatics, and Decision Systems Engineering at Arizona State University (ASU). His Ph.D. research involved designing energy-efficient VLSI systems, accelerating compressive sensing signal reconstruction, and developing emerging memory technology. His current research interests are focused on algorithm, hardware, and system innovations for data analytics and information processing, with emphasis on bringing energy efficiency and data intelligence into a broad spectrum of today’s computing infrastructures, from data center server systems to wearable and Internet-of-things devices. He is a member of the Digital Signal Processing Technical Committee and VLSI Systems \& Applications Technical Committee of the IEEE Circuits and Systems Society.

Dr. Ren received the Broadcom Fellowship in 2012, the prestigious National Science Foundation (NSF) Faculty Early Career Development (CAREER) Award in 2017, the Google Faculty Research Award in 2018. He also received the Top 5 percent Best Teacher Awards from the Fulton Schools of Engineering at ASU in 2017, 2018, and 2019.
\end{IEEEbiography}




\end{document}